\documentclass[preprint,12pt]{elsarticle} 

\usepackage{graphicx}
\usepackage{float}
\usepackage{subcaption}
\usepackage{changepage}
\usepackage{booktabs}
\usepackage{algorithm}
\usepackage{algorithmic}
\usepackage{enumerate}
\usepackage[T1]{fontenc}
\usepackage{amssymb}
\usepackage{pdflscape}
\usepackage{afterpage}
\usepackage{capt-of}
\usepackage{multirow}
\usepackage{array}
\usepackage[table]{xcolor}
\usepackage{algorithm}
\usepackage{algorithmic}
\usepackage{enumerate}
\usepackage{setspace}
\usepackage{todonotes}
\usepackage{amsmath}
\usepackage{comment}
\usepackage{changepage}
\usepackage{upgreek}
\usepackage[normalem]{ulem}
\usepackage{textcomp}
\usepackage{atbegshi,ifthen,listings,tikz}
\tikzstyle{highlighter} = [
  gray!25,
  line width = 0.6\baselineskip,
]

\newcounter{highlight}[page]

\AtBeginShipout{\AtBeginShipoutUpperLeft{\ifthenelse{\value{highlight} > 0}{\tikz[remember picture, overlay]{\foreach \stroke in {1,...,\arabic{highlight}} \draw[highlighter] (begin highlight \stroke) -- (end highlight \stroke);}}{}}}

\newcommand{\minitab}[2][l]{\begin{tabular}{#1}#2\end{tabular}}
\usepackage{amssymb} 

\newcommand{\systemname}{GEESE}

\def\isfinal{0}
\newcommand{\rev}[1]{\if \isfinal0{\textcolor{blue}{#1}}\else {#1}\fi}

\journal{.}
\usepackage{enumitem} 
\setlist{nolistsep}

\let\OLDthebibliography\thebibliography
\renewcommand\thebibliography[1]{
  \OLDthebibliography{#1}
  \setlength{\parskip}{0pt}
  \setlength{\itemsep}{0pt plus 0.3ex}
}

\begin{document}

\begin{frontmatter}

\title{Edge Computing Enabled by Unmanned Autonomous Vehicles}

\author[rvt1]{Mohan Liyanage
}
\ead{mohan.liyanage@ut.ee}

\author[rvt1]{Farooq Dar
}
\ead{farooq.dar@ut.ee}

\author[rvt1]{Rajesh Sharma
}
\ead{rajesh.sharma@ut.ee}

\author[rvt1]{Huber Flores
}

\ead{huber.flores@ut.ee}

\address[rvt1]{Institute of Computer Science, \\ University of Tartu, Estonia, \\
firstname.lastname@ut.ee}

\newcommand\bparagraph[1]{\vspace{0.5mm}\noindent\textit{#1:}}
\newcommand\cparagraph[1]{\vspace{0.5mm}\noindent\textbf{#1:}}

\begin{abstract}

Pervasive applications are revolutionizing the perception that users have towards the environment. Indeed, pervasive applications perform resource-intensive computations over large amounts of stream sensor data collected from multiple sources. This allows applications to provide richer and deep insights into the natural characteristics that govern everything that surrounds us. A key limitation of these applications is that they have high energy footprints, which in turn hampers the quality of experience of users. While cloud and edge computing solutions can be applied to alleviate the problem, these solutions are hard to adopt in existing architecture and far from become ubiquitous. Fortunately, cloudlets are becoming portable enough, such that they can be transported and integrated into any environment easily and dynamically. In this article, we investigate how cloudlets can be transported by unmanned autonomous vehicles (UAV)s to provide computation support on the edge. Based on our study, we develop~\systemname, a novel UAV-based system that enables the dynamic deployment of an edge computing infrastructure through the cooperation of multiple UAVs carrying cloudlets. By using~\systemname, we conduct rigorous experiments to analyze the effort to deliver cloudlets using aerial, ground, and underwater UAVs. Our results indicate that UAVs can work in a cooperative manner to enable edge computing in the wild.

\end{abstract}

 
\end{frontmatter}

\section{Introduction} \label{sec:introduction}

Pervasive sensing is emerging as a fundamental field of study that is transforming the perception that users have towards their environment. Thanks to the proliferation of embedded sensors, wearable, and IoT devices, along with the increasing computing capabilities of smartphones, it is possible to enable a new type of mobile applications that can provide richer and deeper insights about the natural characteristics that govern everything that surrounds us. Examples of this include, analyzing the properties of liquids with cameras~\cite{yue2019liquid}, understanding the quality of food with WiFi signals~\cite{ha2018learning}, identification of unlabeled medicine with infrared optical sensors~\cite{klakegg2018assisted}, spotting hidden cameras with thermal footprints~\cite{flores2020estimating} and pedestrian detection for navigation of autonomous cars~\cite{luckow2015automotive}. One key limitation of this type of applications is that they require analyzing long streams of sensing data, which leads to high energy consumption that drains the battery of end-devices. This is problematic as it hampers the quality of experience (QoE) of users with applications significantly~\cite{ickin2012factors}. To overcome this problem, smartphone applications can offload heavy computations to the cloud and edge servers~\cite{flores2015beyond}, such that devices can reduce energy consumption~\cite{flores2017evidence}. Unfortunately, cloud and edge computing solutions are not well suited to offload long streams of sensing data. Cloud computing is difficult to use due to oscillating communication latency to remote servers, which can induce additional energy overheads and slow down the performance of applications. Likewise, edge computing is not available in an ubiquitous manner. Indeed, edge deployments are scarce and not dense enough to provide stable services. Nevertheless, edge computing solutions are preferable over cloud computing ones as edge computing can analyze long streams of sensor data with consistent performance, and without inducing high energy overheads in end-devices caused by moving sensor data too far from its source~\cite{satyanarayanan2015edge}.

To overcome the scarcity of edge computing deployments, most of the existing work has focused on exploring two prominent perspectives. The first one explores the formation of underlying computing infrastructure from resources of devices located densely in the environment, e.g., personal computers, smart and IoT devices, such that edge services can be executed on top of it~\cite{lagerspetz2019pervasive}. The other perspective consists in identifying the optimal placement of static edge servers~\cite{wang2019edge}, such that the coverage of the edge computing support can be ensured in a particular area (see Section~\ref{sec:relatedwork} to obtain detailed description). While these solutions can partially alleviate the ubiquitous access to edge computing resources, they do not tackle the problem of scalability of edge computing. Indeed, unlike cloud computing that is elastic, which means that computational resources can be added or removed dynamically to handle the dynamic workload of users~\cite{herbst2013elasticity} (a number of concurrent users), edge computing completely lacks the properties of elasticity. Consequently, edge resources have limited capacity and are unable to handle dynamic workloads. For instance, consider a large crowd of people that gather spontaneously in a location, e.g., flash mobs~\cite{nicholson2005flash}. If a single edge server is available to process streams of sensing data for the crowd, then the server can easily be overloaded by such a group, and instead of improving the QoE of users, it can end up degrading the overall performance of all the devices in the crowd. Fortunately, cloudlets~\cite{satyanarayanan2009case} are becoming portable and they are the fundamental source for providing computing support on the edge. Cloudlets can be easily created from devices that have small sizes and low weights. For instance, it has been demonstrated that smartphones can be used as cloudlets~\cite{habak2015femto, lagerspetz2019pervasive}. Thanks to this portability, it is possible then to envision cloudlets that can be transported and integrated into any environment dynamically. Recent advancements in unmanned autonomous vehicles (UAV), which are also known as drones, can provide transportation means to carry cloudlets to the edge.

In this paper, we contribute by developing~\systemname\footnote{GEESE refers to a group of social birds that are adapted to ground, water and aerial environments.}, a novel UAV-based system that enables the dynamic deployment of an edge computing infrastructure though cooperation of multiple UAVs carrying cloudlets. Since different UAV transportation modalities can be used to transport cloudlets, we study the transportation of cloudlets using aerial, ground, and underwater UAVs (Here thereafter we will refer to any drone modality as UAV). To this end, we explore the requirements to be fulfilled for equipping different types of UAVs with cloudlets, and the effort required by UAVs to move cloudlets of different sizes and weights between locations. In addition, we also analyze the cooperation between UAVs to form an edge computing infrastructure via collaborative processing. To scale up cloudlets on the edge, we also explore a resource delivery model that enables us to optimally estimate the amount of UAVs and cloudlets required to handle a workload of users on the edge. Our results with~\systemname~demonstrate that by working in a collaborative manner, UAVs can transport cloudlets to enable the dynamic deployment of an edge computing infrastructure close to users.

\hfill \break
\noindent The contributions of the paper are summarized as follows:
\hfill \break
\begin{itemize}

\item \textbf{New insights about using UAVs to enable edge computing deployments with cloudlets:} We quantify the performance of UAVs to transport cloudlets to the edge. We show that weight and the encasing of the cloudlet are critical factors to equip different types of UAVs with cloudlets.

\item \textbf{Improved and scalable edge computing deployments:} We demonstrate that an edge computing infrastructure can be formed through the cooperation of multiple types of UAVs via collaborative processing. 

\item \textbf{New method to transport and place cloudlets on the edge dynamically:} We develop an edge delivery model that estimates the amount of UAVs and capabilities of cloudlets required to satisfy a workload from a crowd of users. In addition, we also demonstrate that UAVs can be easily used to reallocate edge computing resources based on the characteristics of users' mobility.
\end{itemize}

\section{Background and related work} \label{sec:relatedwork}

In this section, we discuss existing work and highlight how our contributions advance the state of the art. 

\noindent\textbf{Cloudlets:} A Cloudlet is a resourceful and trusted piece of computing infrastructure located close to end users~\cite{satyanarayanan2009case}. Cloudlets are among the first architectural models that emerged for providing computing support on the edge, e.g., cyber-foraging~\cite{balan2002case}. Cloudlets can exploit proximal and static infrastructure, such as hotspots and base stations, in an opportunistic manner to push computing services into networks that are accessible by devices of end-users through low and consistent latency. The main problem with cloudlets is that it requires the deployment of dedicated infrastructure, e.g., servers, which is difficult to maintain and deploy permanently in a location. In addition, as cloudlets are accessed through proximity-based communications e.g., WiFi-Direct and Bluetooth, the computing support coverage of cloudlets in an area is limited to a few hundred meters or less as it depends on the characteristics of the environment where the cloudlet is deployed, e.g., indoor building, outdoor park. As a result, the success -- or failure-- of a cloudlet deployment depends on a dense distribution of a high number of cloudlets in a location. Dense deployment of cloudlets in the wild is critical for applications, such that they can operate transparently, without interruptions nor degradation in the quality of service (QoS). Unlike previous work, we do not focus on the static deployment of cloudlets, but instead, we focus on the dynamic deployment of cloudlets using UAVs.

\noindent\textbf{Mobile, transient and opportunistic computing infrastructures:} To overcome the scarce distribution of cloudlets, smart devices have been investigated to form a mobile infrastructure to provide computing support on the edge, for instance, a cluster of smartphones~\cite{busching2012droidcluster}. Indeed, since the computational capabilities of smart devices have increased dramatically, it is possible to provision cloudlets through smart devices~\cite{habak2015femto, lagerspetz2019pervasive}. Moreover, smart devices can exploit the characteristics of human mobility to move between locations in a transient manner -- similar concept has been adopted by Microsoft Data Box Edge. Thanks to this mobility, multiple devices also can be co-located opportunistically in a proximal communication range of each other~\cite{conti2010opportunities}. As a result, collaborating computing scenarios between devices (mobile crowd) can be envisioned to form an aggregated mobile computing infrastructure~\cite{flores2017social, zhang2018towards}. This also includes IoT devices that are distributed across the environment~\cite{koukoumidis2011pocket, verbelen2012cloudlets}. In this type of infrastructure, multiple devices merge their individual and constrained computing resources into a single pool that has higher computing capabilities. A mobile infrastructure can provision (cloudlet) services on its own or augment static edge and fog deployments~\cite{satyanarayanan2017emergence, su2018distribution}. While mobile infrastructures are feasible, in practice, it raises many privacy concerns from users' perspective~\cite{flores2017social}. To use a mobile infrastructure, several frameworks to distribute computational tasks among multiple devices have been proposed over the years~\cite{dou2010misco, marozzo2012p2p, guo2018foggycache}. For instance, frameworks for collaborative sensing and computing have been proposed~\cite{amiri2014rio, lee2012comon, flores2020cosine}, as well as frameworks to distribute machine and deep learning tasks among multiple devices to achieve edge intelligence~\cite{lagerspetz2019pervasive, li2018learning}.  Unlike other work, we focus on transporting cloudlets with UAVs. These cloudlets can be then used to form a mobile infrastructure (micro-data centres) through collaborative processing among UAVs.

\noindent\textbf{Edge computing placement:} Edge computing deployments that are closer to end-devices are emerging with the promise to provide better QoS and QoE for end applications. Edge computing enables the processing of big data rapidly without moving it too far from its source, which in turn improves the performance of end-devices and applications. Unfortunately, the deployment of permanent and specialized hardware, e.g., servers, is required to provide such computing support on the edge. Since the deployment and maintenance of edge infrastructure has a cost and its capacity to handle computing workload is limited (a number of concurrent users), its deployment optimization and resource allocation is an important issue to solve for cloud and mobile operator vendors~\cite{brimberg2008survey, wang2020flat}. Consequently, the placement of edge servers has been explored, taking into consideration multiple optimization trade-offs~\cite{wang2019edge, sharma2016multi}. The allocation of cloudlets as edge servers has been analyzed given a budget constraint~\cite{fan2017cost}. Trade-offs between the number of servers and deployment cost while providing efficient QoS and QoE also have been investigated~\cite{sinky2019adaptive, mondal2018ccompassion, lee2019low}. Other trade-offs maximize the deployment of servers in a geographical area to ensure edge service coverage~\cite{yin2016edge, jia2015optimal}. Several other works focus on minimizing latency to edge servers while handling different dynamic workloads of users~\cite{yin2016edge}. The placement of edge servers to reduce the energy consumption of devices also has been studied~\cite{li2018energy}. Unlike other works that focus on optimizing static deployments, we focus on equipping UAVs with cloudlets to transport computation support to the edge dynamically. In this manner, edge computing resources can be placed temporally in a location based on the demand of users and then re-allocated easily based on the characteristic of users' mobility.

\noindent\textbf{Delivery and cooperation of UAVs:} Applications for UAVs (also known as drones) are broad and range from autonomous explorations to area mapping and monitoring~\cite{motlagh2016low, nonami2010autonomous}. The idea of using UAVs for the delivery of goods and services have been explored from commercial~\cite{prager2018payload} and scientific~\cite{prager2018payload} perspectives. Among goods and services delivered by aerial drones, the delivery of sensitive information~\cite{yoon2017adaptive}, networking resources to form a network in emergency situations~\cite{zhang2017optimal}, and a single micro-cloud~\cite{sathiaseelan2016cloudrone} have been studied. Several centralized and decentralized architectures for UAVs have been proposed over the years~\cite{jiang2018laair, paull2014decentralized}. Optimization of trajectories and autonomous navigation has been explored in detail using different UAV transportation modalities~\cite{wang2017autonomous}, including ground~\cite{vandapel2006unmanned}, aerial~\cite{sinopoli2001vision, kurnaz2009fuzzy} and underwater~\cite{ermayanti2015estimate} modalities. In terms of cooperation between UAVs, models have been designed to achieve common goals through cooperation~\cite{datta2017vehicles, loke2019cooperative}, this includes swarm~\cite{zhang2016social}, clustering~\cite{tazibt2017wireless} and social~\cite{butt2018social} cooperation models. Localization systems for UAVs to achieve cooperative behavior also have been proposed~\cite{bahr2009cooperative}, e.g., sampling grids~\cite{saeed2017argus}. Unlike previous works, we focus on equipping different UAV modalities with cloudlets. We envision cooperation between multiple UAVs to create dynamic edge computing infrastructure. We also focus on the delivery and re-allocation of computing resources on the edge based and on the workload demand of mobile users.

\noindent\textbf{Summary and differences to previous work:} To summarize, pushing cloudlets to the edge to provision services have been explored. Cloudlets in static infrastructure, e.g., hotspots, have been envisioned in traditional architectures, and many strategies to place static servers have been developed. Smart and IoT devices to form mobile infrastructures to deploy cloudlets also have been investigated. UAVs to deliver cloudlets have been just explored for aerial UAVs. In this work, we explore the transportation performance of computing resources to the edge using aerial, ground, and underwater modalities. We explore the necessary encasing that needs to be considered for each modality. We also explore the cooperation between multiple UAVs to form a computing infrastructure on the edge (micro-data center). In addition, we also study the deployment of edge computing resources to handle the dynamic workload and its re-allocation based on the mobility demand of users.

\section{Portable cloudlets: Feasibility analysis}

The focus of our work is to investigate the usage of UAVs to transport cloudlets to the edge. Before detailing our methodology, we investigate a fundamental question to equip UAVs with cloudlets. We examine the feasibility of using smart and IoT devices as cloudlets. While previous work has demonstrated that these devices have enough resources to execute cloudlet services smoothly~\cite{habak2015femto, lagerspetz2019pervasive}, we analyze further the capacity of smart and IoT devices to handle the different workload of users when acting as cloudlets. This information is critical to determine whether portable cloudlets can provision computing services to a large number of users. In addition, since the type of application that is executed by the cloudlet also plays a significant role in the provisioning of services, we also analyze the performance of different types of applications when running in portable cloudlets.

\subsection{Computing capacity and energy consumption of cloudlets}

Since UAVs need to carry a cloudlet from a source location to a target destination, it is critical to minimize payload weight and size, while at the same time maximizing the amount of computation capacity that is delivered. Increased computational capabilities and reduced sizes of smart and IoT devices make them suitable candidates to create cloudlets that can be embedded in UAVs. To thoroughly characterize the computational capacity of smart and IoT devices, we conduct rigorous benchmarks that evaluate the influence of the increasing workload of users on the throughput of the devices. Moreover, we also analyze the influence of constant workload on the battery of the resources of devices acting as cloudlets. In our experiments, we analyze the computational capacity to handle the workload of two different devices, a Samsung Galaxy S5 (S5) and a Raspberry Pi 4 Model B (RP4). These devices are selected as the former generalizes computing resources that can be found in common smartphones, whereas the latter is used widely in most IoT prototype solutions. We transform these devices into cloudlets that provide services that can be requested by other users following a client/server architecture. As a task running by the service, we design a computing task that provides a constant and uniform response time and also depicts resource-intensive processing. By controlling these properties of the task, we ensured that the benchmark is not influenced by non-deterministic computing behavior. The selected task consists of detecting prime numbers within a list of available numbers. In this context, a client sends a request to the cloudlet service. As part of the request, the client sends a list of $20$ integer numbers, which are within the range of $100000-105000$. The cloudlet service takes the request and identifies the prime numbers in the list, and sends the result back to the client. To analyze multiple clients sending requests to the service, we use JMeter\footnote{https://jmeter.apache.org/} to simulate different workloads of users. We use an increasing workload from $1$ to $100$ users to analyze the capacity of the cloudlets to handle the workload.

\begin{figure*}[htb]
\centering
  \begin{subfigure}[b]{.5\linewidth}
    \centering
    \includegraphics[width=1.0\textwidth]{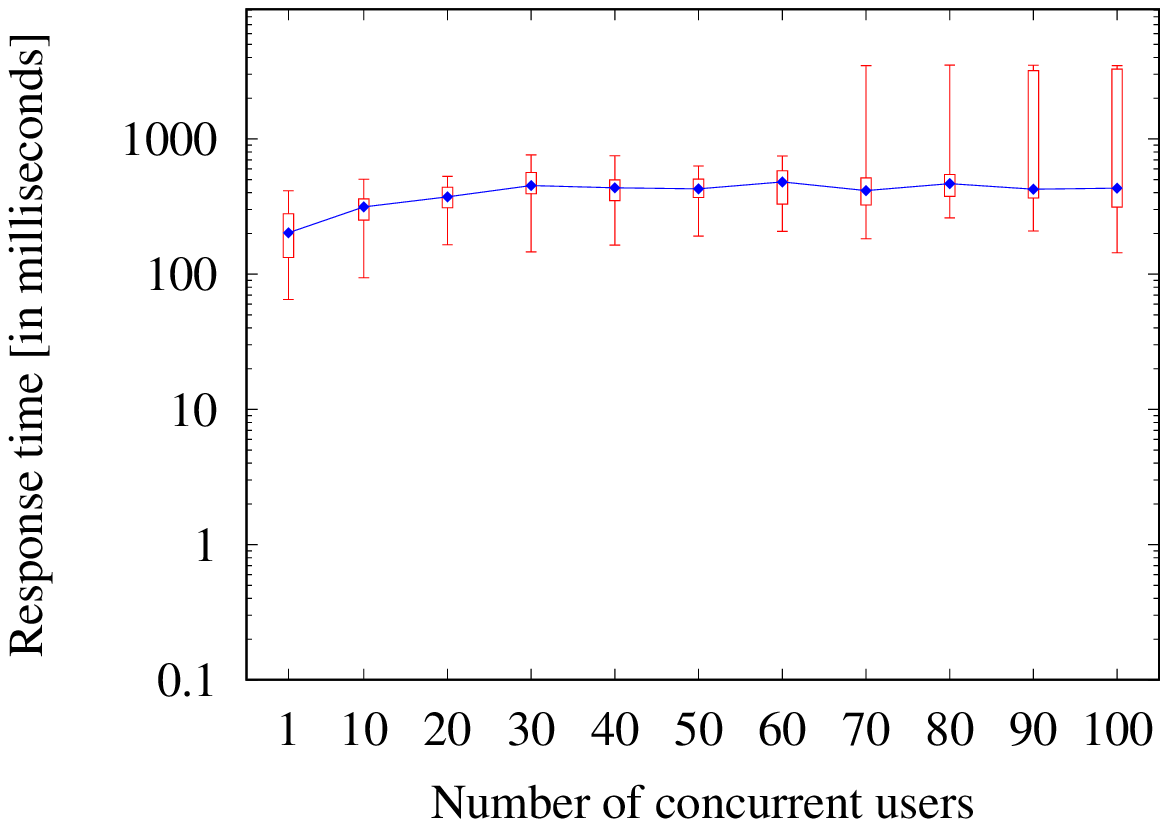}
   \caption{Performance of S5 to handle workloads.}\label{fig:smartphoneload}
  \end{subfigure}%
  \begin{subfigure}[b]{.5\linewidth}
    \centering
    \includegraphics[width=1.0\textwidth]{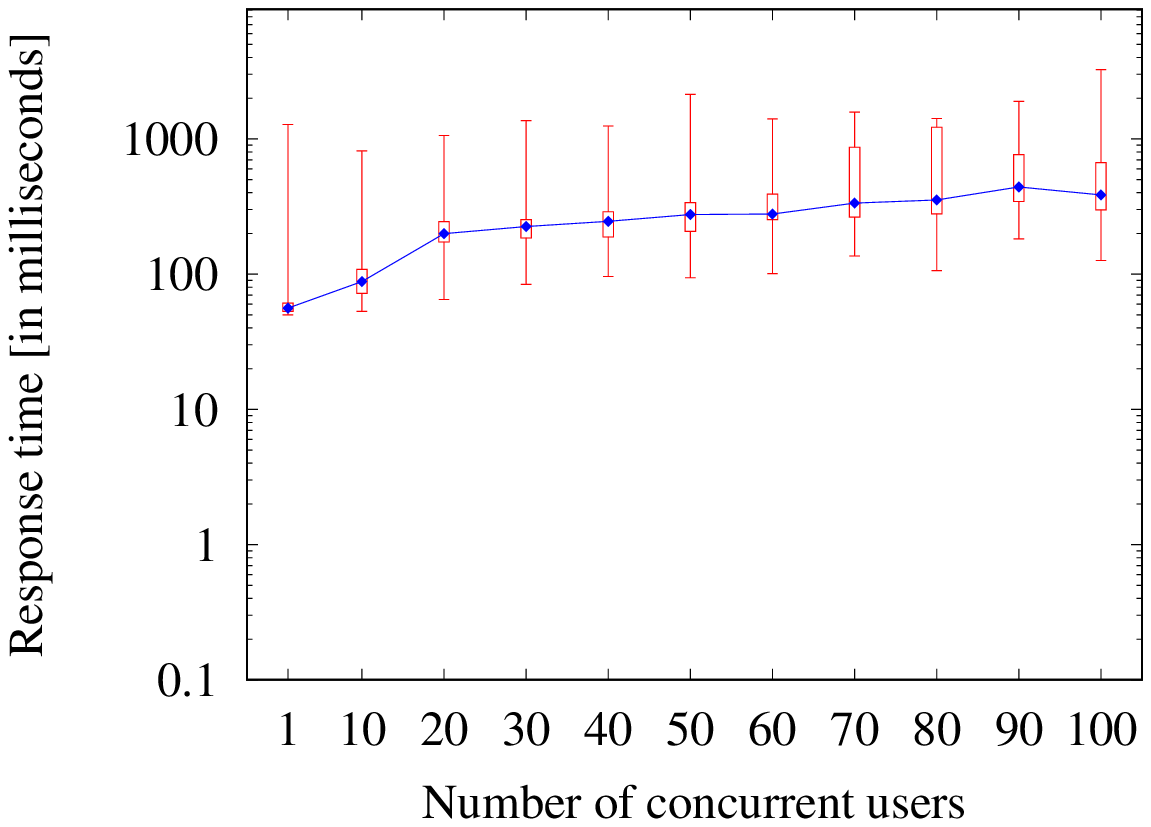}
    \caption{Performance of RP4 to handle workloads.}\label{fig:raspberyyload}
  \end{subfigure} \\
  \begin{subfigure}[b]{.5\linewidth}
    \centering
    \includegraphics[width=1.0\textwidth]{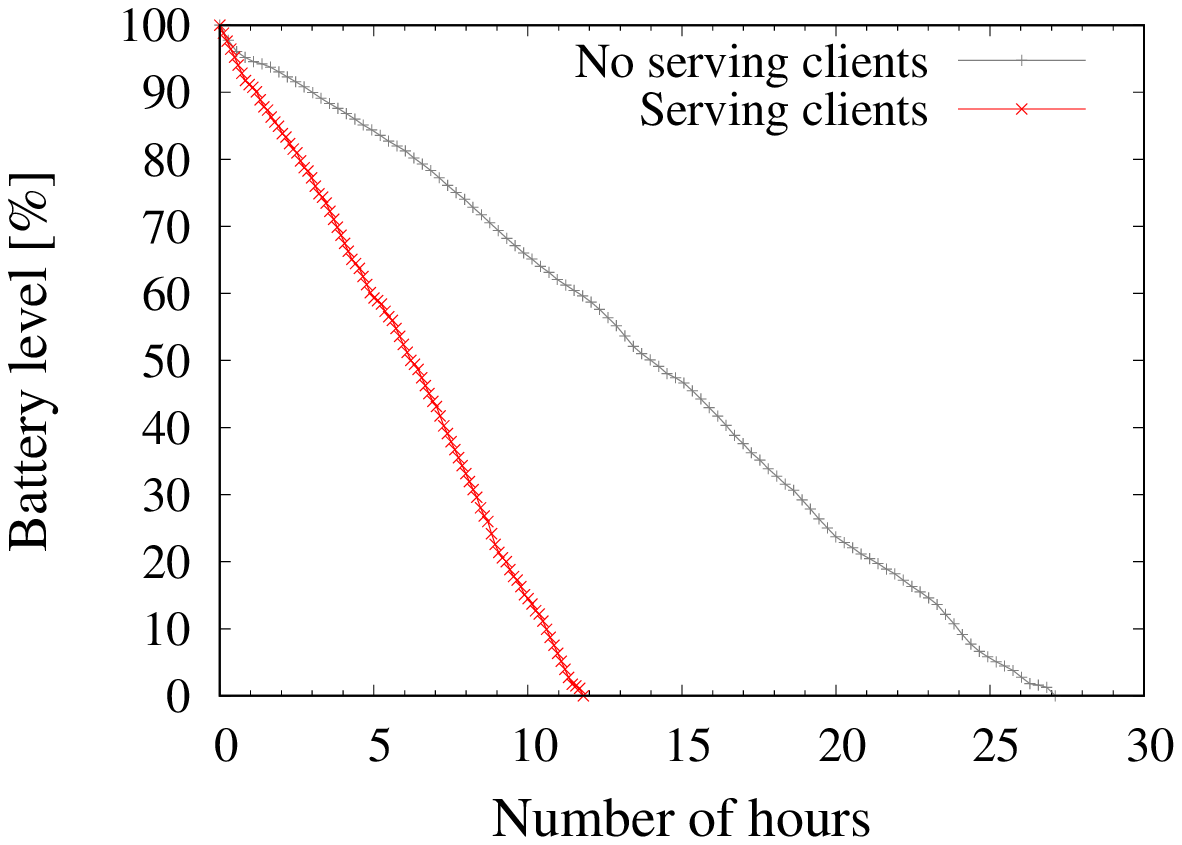}
   \caption{Energy consumption of S5}\label{fig:smartphoneenergy}
  \end{subfigure}%
  \begin{subfigure}[b]{.5\linewidth}
    \centering
    \includegraphics[width=1.0\textwidth]{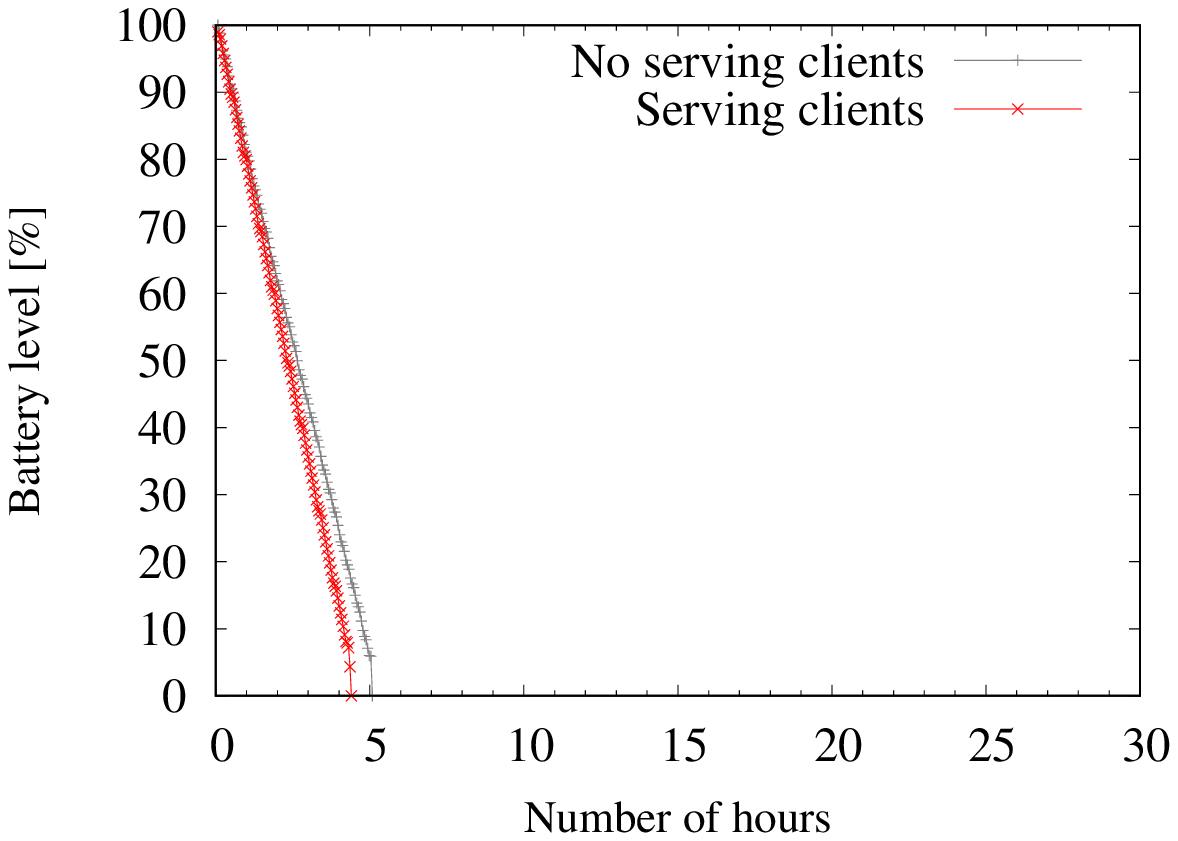}
    \caption{Energy consumption of RP4}\label{fig:raspberyenergy}
  \end{subfigure}%
  \caption{Capacity of cloudlets powered by smart and IoT devices. a) Number of concurrent users handled by a Samsung S5, b) Number of concurrent users handled by a RP4. Battery drain of devices when facing a fixed workload of $100$ concurrent users, c) Battery drain for S5, d) Battery drain for RP4.}
  \label{fig:DevicesLoad}
\end{figure*}

Figure~\ref{fig:DevicesLoad} shows the results of the experiment for both cloudlet S5, and cloudlet RP4. In both cases, we can observe that the response time of the requests increases with an increase in the number of concurrent requests. In addition, we also note that S5 provides more constant response time when handling the same workload when compared with RP4. We can observe this as the interquartile range of RP4 has a slightly higher slope than S5. In terms of a concurrent number of users, we can observe that Figure~\ref{fig:DevicesLoad}a indicates that S5 can handle workloads up to $100$ concurrent users, while still providing response time in the range of milliseconds. Similarly, we can observe Figure~\ref{fig:DevicesLoad}b, it can be noted that RP4 can handle workloads up to $100$ concurrent users with better average response time than S5. In both cases, we can appreciate a reduction in throughput as load increases, which also impacts the overall response time of requests. For instance, we can observe that the response time increases $\approx 3$ times when workload increases from $20$ to $90$ concurrent requests (response time $371$ ms and $1149$ ms respectively) for S5. Similarly, we can observe that response time increases $\approx 2.5$ times when the workload is increased for RP4 (from $20$ to $90$ concurrent users).

Additionally, since the battery resources of smart and IoT devices are limited, we analyze how long these resources last when experiencing a fixed workload of users. We select a fixed workload of $100$ concurrent users for this experiment as for both cases, this is the maximum number of users that can be handled by the cloudlets, while still keeping the response time in milliseconds. With this workload, all the sent requests were completed and cloudlets did not crash due to over load. Figure~\ref{fig:DevicesLoad} also shows the result of the energy drain for both devices. From the results, we can observe that the battery of S5 can handle a workload of $100$ users for $11$ hours (Figure~\ref{fig:DevicesLoad}c), while RP4 can handle the same workload during just four hours (Figure~\ref{fig:DevicesLoad}d).

\noindent \textbf{Insights:} Our results demonstrate that smart and IoT devices can be used as cloudlets to provide computing services to users in the wild. The results indicate that a single device unit can handle up to $100$ concurrent users for at least four hours in the minimum case. This suggests that small and low weight devices can be integrated dynamically into any environment to overcome the scarcity of edge infrastructure when required. Naturally, the type of application is critical when using portable cloudlets as underlying computing infrastructure. Computing power is exploited differently based on application type. In this use case, web services rely on CPU for processing, however, other applications, such as image processing, can exploit GPU hardware acceleration to improve performance (See Section~\ref{sec:resultsgeese} for further details about running image processing in portable cloudlets).

\subsection{Application performance on cloudlets}

As demonstrated previously, portable cloudlets made from smart and IoT devices can provide services to a large group of users on the edge. However, another concern in doing this, it's the quality of service and perception that users have towards using the applications run by the cloudlets. Indeed, different applications have different computational requirements for providing high levels of responsiveness and performance. For instance, the amount of computing resources required by deep learning applications are higher when compared with applications that implement simple linear regression routines~\cite{li2018learning}. Thus, cloudlets also have to be selected based on the processing requirements of applications. On the other hand, to overcome the limitations of individual cloudlets, it is possible to rely on collaborative processing dynamically~\cite{FemtoCloud, lagerspetz2019pervasive, flores2020cosine}. In this case, the execution of an application is distributed among multiple devices. Resource fragmentation plays a major role when relying on distributed devices. While it is possible to improve performance by aggregating computing resources dynamically~\cite{amiri2014rio}, computing resources whose processing capabilities differ significantly can easily become a bottleneck for applications. This mismatch in processing capabilities can cause applications to slow down their performance and ultimately make them not usable.

\begin{figure*}[htb]
    \centering%
     \begin{subfigure}[b]{0.5\columnwidth}
    \centering
        \includegraphics[width=.90\columnwidth]{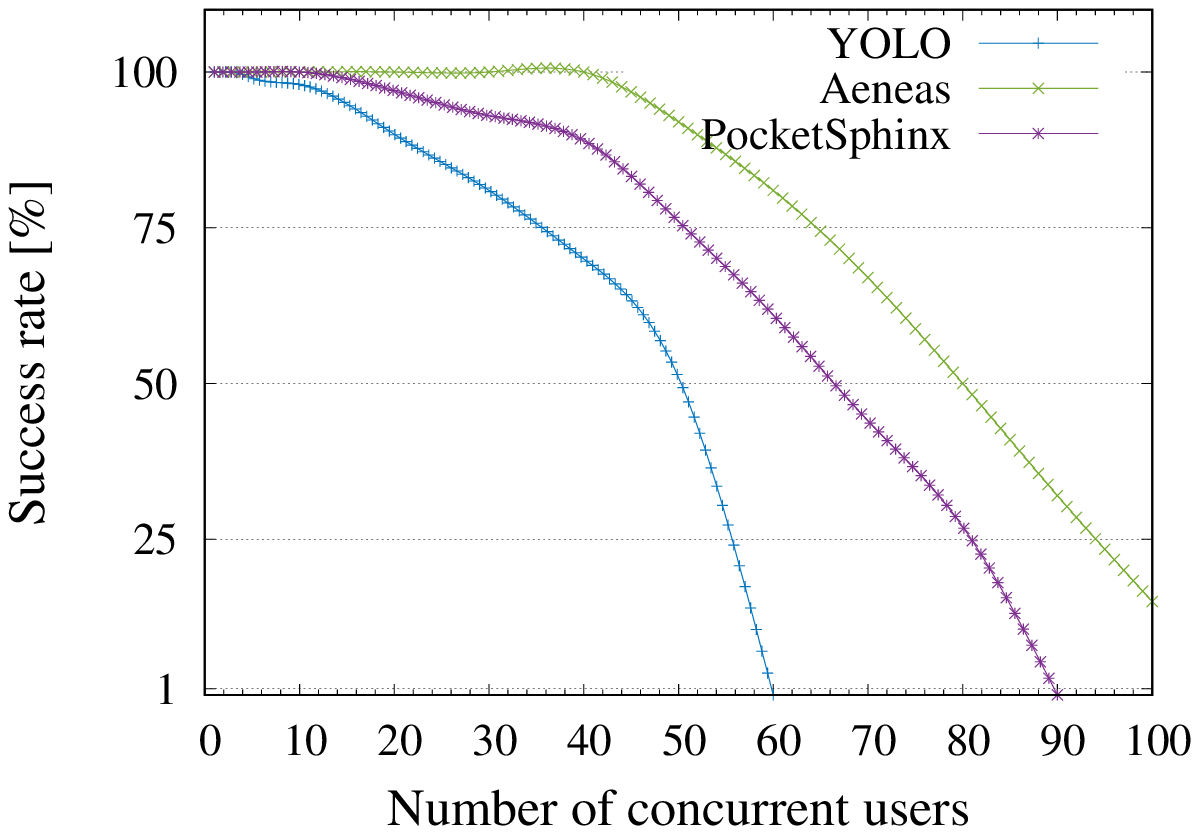} 
        \label{fig:cloud1}
       \caption{Sucess rate}
    \end{subfigure}%
     \begin{subfigure}[b]{0.5\columnwidth}
    \centering
        \includegraphics[width=.90\columnwidth]{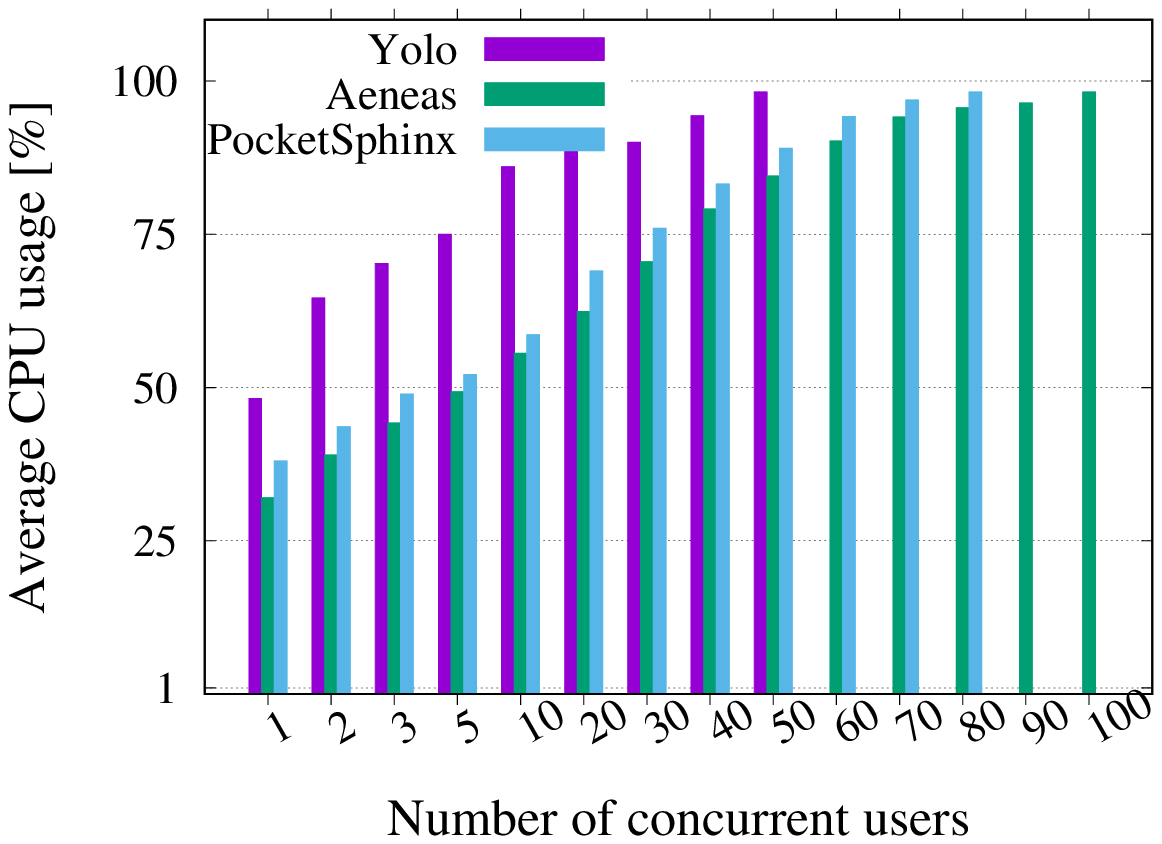}
        \label{fig:edge1}
        \caption{Average CPU usage}
    \end{subfigure} \\
    \begin{subfigure}[b]{0.5\columnwidth}
    \centering
        \includegraphics[width=.90\columnwidth]{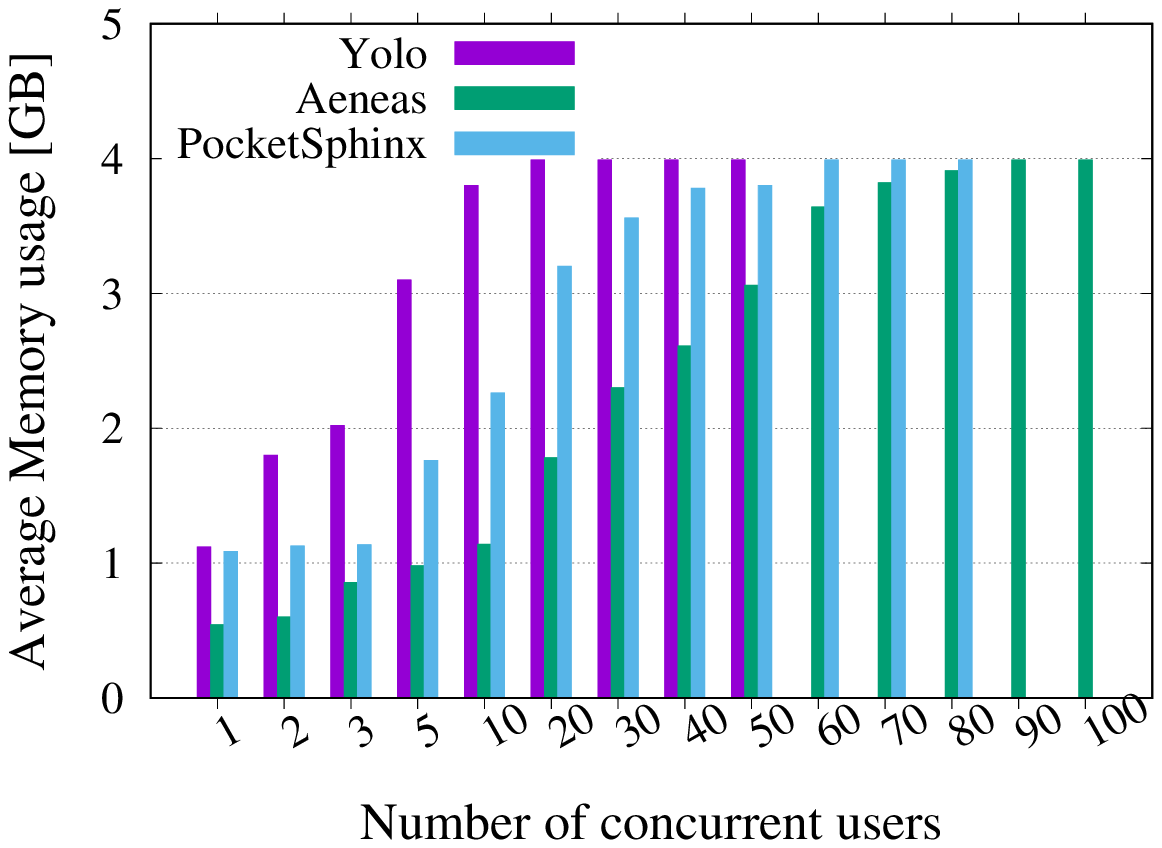}
        \label{fig:uavs1}
         \caption{Average memory usage}
    \end{subfigure}%
    \caption{Benchmark analysis of applications run on RP4}
     \label{fig:AppPerformance}
\end{figure*} 

To estimate the processing requirements of applications, a general solution is to rely on computing benchmarks~\cite{shukla2017riotbench, mcchesney2019defog}. Thus, to analyze the performance of applications run by portable cloudlets, we conduct further experiments using the DeFog~\cite{mcchesney2019defog} platform. DeFog provides a set of applications to benchmark edge computing deployments, and provides a diversity of applications to measure performance. We analyze the performance of different types of applications when running in our RP4 cloudlet. As applications, we relied on YOLO (Deep learning for object classification app), PocketSphinx (Speech to text conversion app) and Aeneas (Text and audio synchronization app). Figure~\ref{fig:AppPerformance} shows the results, we can observe that resource intensive applications limit the amount of requests from multiple users. For instance, the success rate of uses requesting YOLO is lower when compared with Aeneas. In addition, we can also observe that the success rate for handling multiple requests starts to drop as the computing resources of RP4 start to become over utilized. This clearly suggests that type of application also plays an essential role when relying on portable cloudlets. We can also appreciate this when analyzing other performance metrics, such as average CPU and memory usage (Figure~\ref{fig:AppPerformance}b and Figure~\ref{fig:AppPerformance}c respectively).

\begin{figure}
\centering
\includegraphics[width=0.75\textwidth]{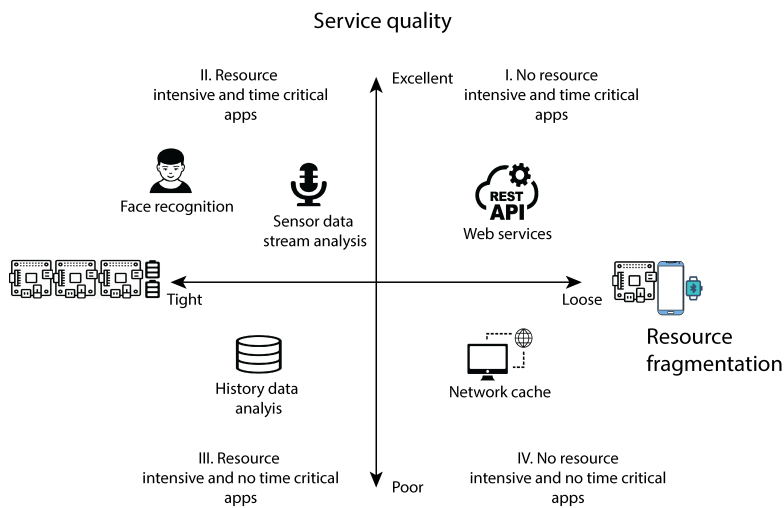}
\caption{Classification of applications running on distributed computing resources.}
\label{fig:app-benchmark}
\end{figure}

\noindent \textbf{Insights:} To model the performance of the application when executed in a distributed manner, we extrapolate the individual application performance of our benchmark from single to multiple devices. To do this, we model application execution depending on the level of resource fragmentation. The level of fragmentation is low when devices share similar processing resources (Tight - all devices are the same), and becomes high when resources are too different (Loose). Our use case of collaborative processing presented in Section~\ref{sec:resultsgeese} shows that aggregated processing resources improves overall performance. In parallel to this, applications experience fewer bottlenecks when resource fragmentation is low. Thus, the service quality of the application is better when executed on tight resources instead of loose ones. Figure~\ref{fig:app-benchmark} provides an overall view of how resource fragmentation and service quality have to be considered when running different types of applications in portable cloudlets. For instance, deep learning applications, e.g., real-time face recognition, require tight resources to provide a high quality of service (Quadrant II). Likewise, when processing requirements are low, e..g, cache data, and response time is not critical (it is asynchronous), resource fragmentation can be then loose (Quadrant IV). Taken together, our results provide insights on how to select underlying computing resources to run different types of applications.
\section{\systemname~design and development}

In the previous section, we demonstrate that portable smart and IoT devices have enough computing capabilities and energy resources to handle heavy computational workloads in the wild. Thus, it is possible to use them as cloudlets. In this section, we introduce~\systemname, a novel UAV-based solution to enable the adaptive transportation and placement of edge computing infrastructure.~\systemname~delivers~\textit{cloudlets on demand} to any environment to provide~\textit{edge computing services} to end-applications. These services include processing resource augmentation, data analytics and computational offloading. In the following, we first reflect on the limitations of existing architectures, and how~\systemname~can help to overcome the existing issues. Next, we give an overview of the design of~\systemname, where we describe all of its components. In addition, we also provide implementation details of our proposed system.

\begin{figure*}
    \centering%
     \begin{subfigure}[b]{0.5\columnwidth}
    \centering
        \includegraphics[width=1.0\columnwidth]{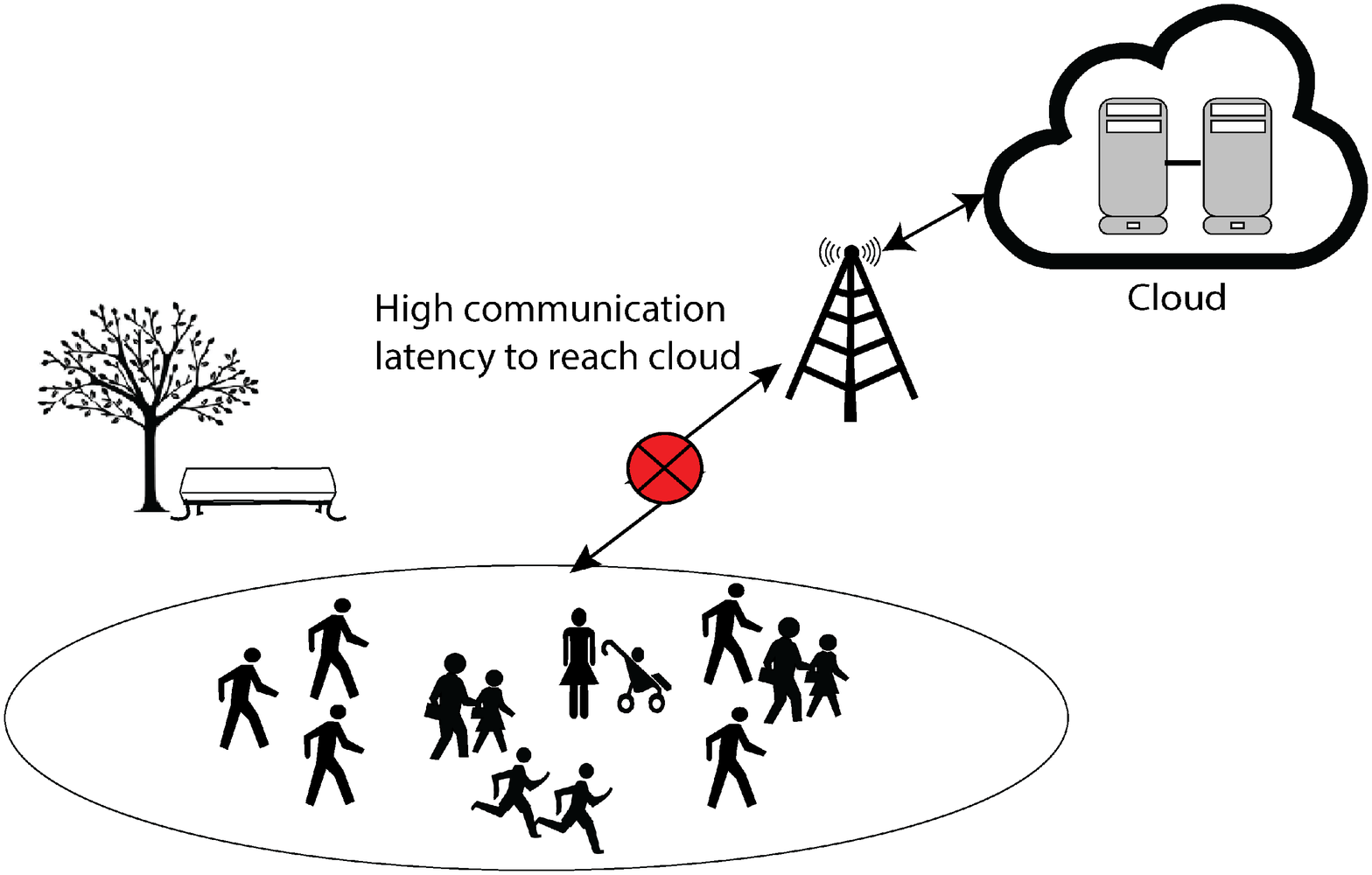} 
        \label{fig:cloud}
        
        \caption{Cloud access}
    \end{subfigure}%
     \begin{subfigure}[b]{0.43\columnwidth}
    \centering
        \includegraphics[width=1.0\columnwidth]{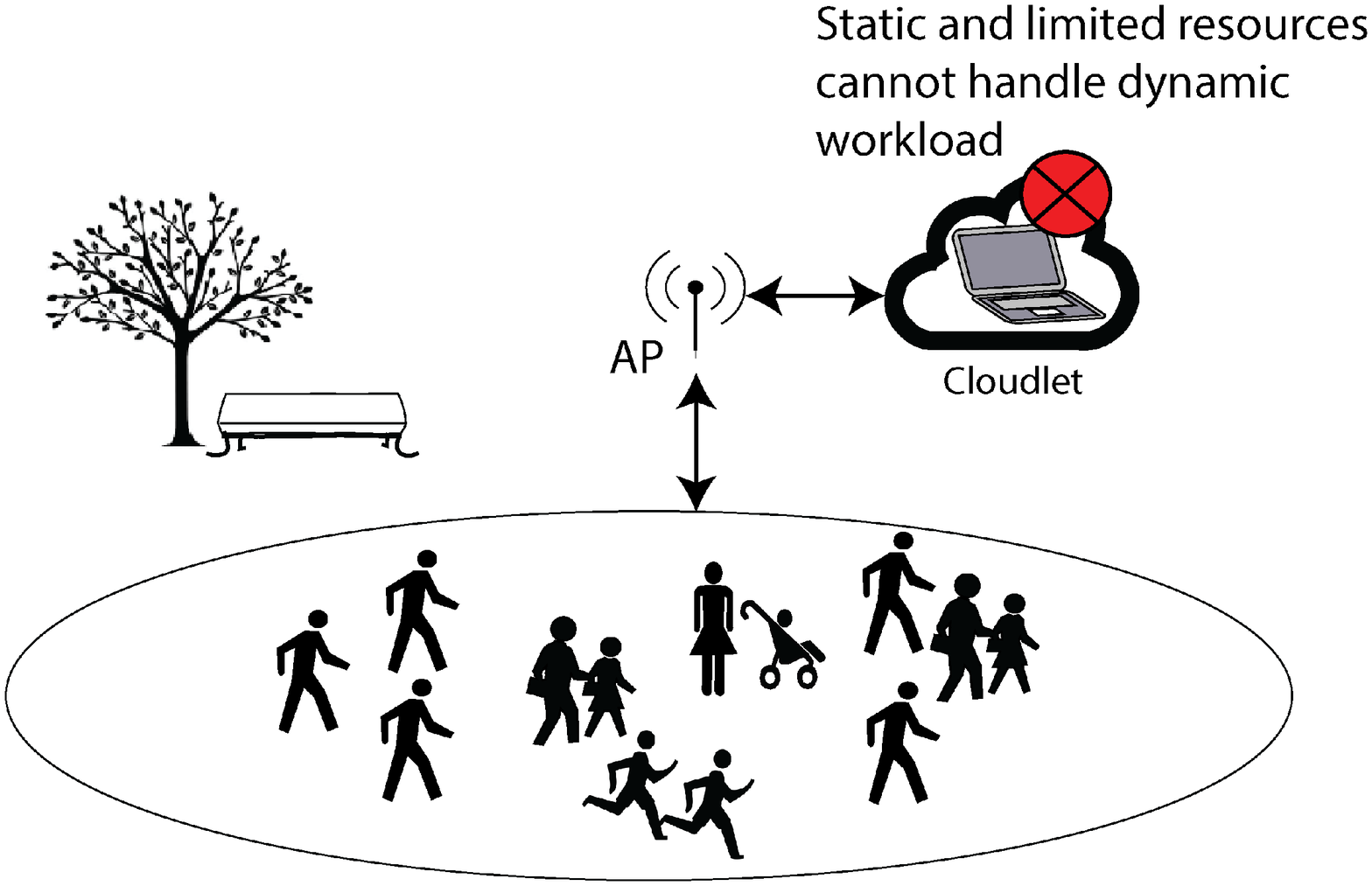}
      
        \label{fig:edge}
       
         \caption{Edge access}
    \end{subfigure} \\
    \begin{subfigure}[b]{0.47\columnwidth}
    \centering
        \includegraphics[width=1.2\columnwidth]{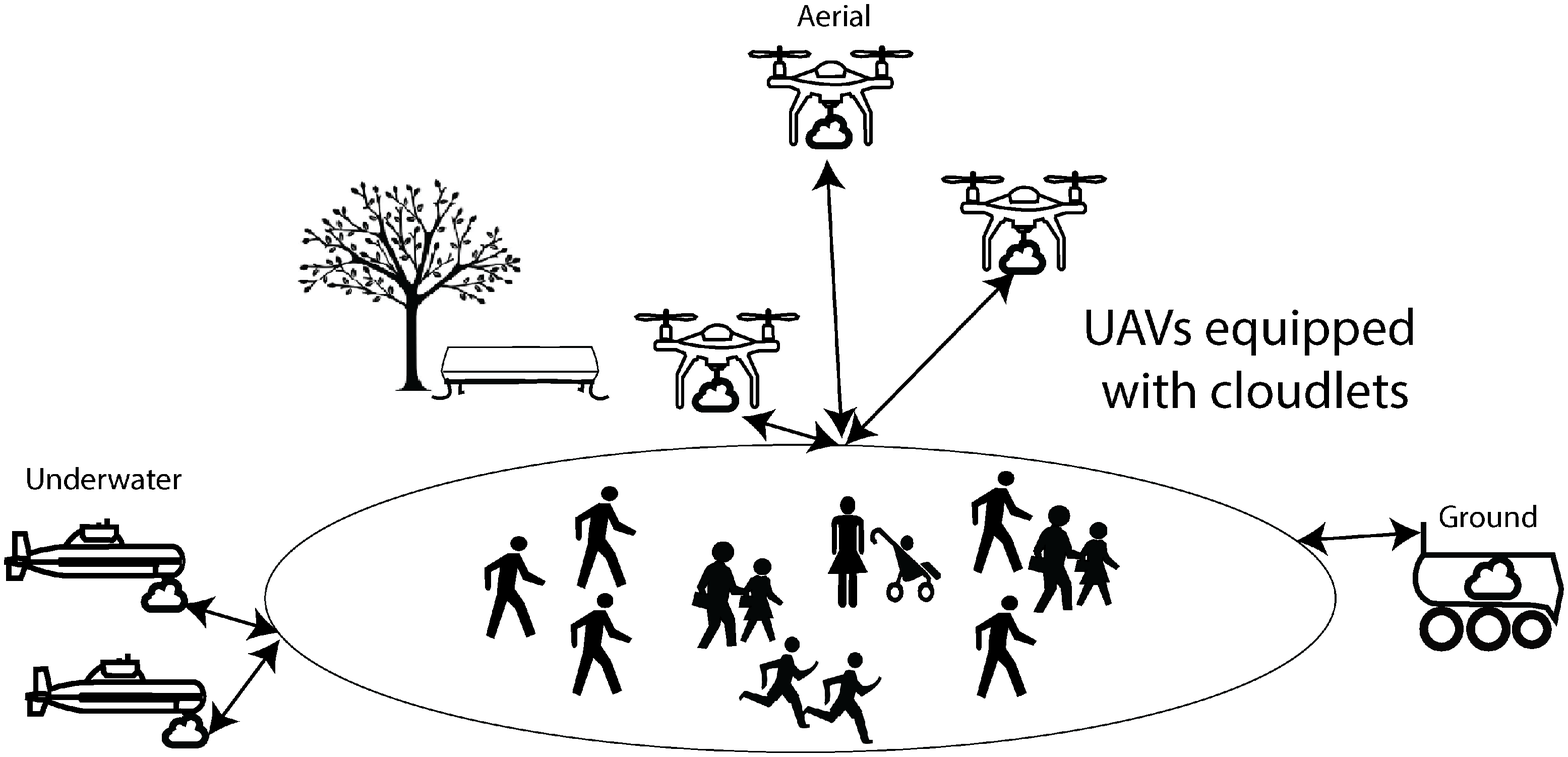}
      
        \label{fig:uavs}
        
         \caption{UAVs access}
    \end{subfigure}%
    \caption{A large group of users looking for computation support on the move, a) Using cellular network to reach remote cloud resources, b) Using Wi-Fi/Bluetooth to access constrained edge server in proximity, c) Using the cloudlet support provided by UAVs in proximity.}
     \label{fig:Overview}
\end{figure*}

\subsection{\systemname~Advantages}

Existing architectures can rely on cloud and edge based solutions to overcome the issues of big data and resource intensive processing in constrained devices. However, these solutions have several limitations, which are related to dynamic users' mobility, oscillating communication latency and availability of deployments. We highlight these limitations in Figure~\ref{fig:Overview}. In the figure, we can observe a large group of users moving together (aka mobile crowd), which is a normal and common situation that can be found in everyday life. For instance, protests and open markets can gather crowds of different sizes~\cite{biggs2018size}. With social media, and a number of pervasive apps available to support and inform about the activities of the crowd, e.g., material recognition and air quality monitoring apps~\cite{motlagh2020toward}, it is necessary to support these devices with external computing services, such that applications can run smoothly while at the same time battery life of devices is preserved. To achieve this, it is possible to rely on cloud and edge computing services while being subject to several drawbacks.

Figure~\ref{fig:Overview}a shows that users can connect to the cloud through the cellular network, however, the transmission performance may be uncertain as remote communication suffers from oscillating changes of latency. For instance, 3G can change drastically from $150$ to $500$ milliseconds delay~\cite{flores2017social}. Moreover, the communication base station where the group is connected can suffer from congestion caused by the large number of users connecting to it. 
Alternatively, edge computing services can be provided to the group as shown in Figure~\ref{fig:Overview}b. These services are deployed close to users, e.g., hotspots, such that they can connect to it directly using low range communication technologies, e.g., WiFi-Direct and Bluetooth. While this approach provides a more consistent quality of service and does not suffer from oscillating latency, it is instead very difficult to find in the wild. Indeed, the availability of edge deployments is not dense enough to be ubiquitous. In addition, edge computing services cannot scale seamlessly due to their limited resources. As a result, edge services can support a partial amount of users in the group as long as they are within the communication range. To overcome these issues, UAVs can be used instead to transport cloudlets to the edge. Figure~\ref{fig:Overview}c shows the overall solution. We can observe that the group connects in direct proximity to cloudlets carried by UAVs. Moreover, UAVs can be scheduled or replaced to support the group continuously, and UAVs can also be re-allocated easily as the group moves. In addition, there is no maintenance and neither deployment cost associated with having fixed servers in a location. In the rest of this section, we describe the design of~\systemname~to transport cloudlets to the edge.

\subsection{\systemname~Design and Components}

\begin{figure}
\centering
\includegraphics[width=0.75\textwidth]{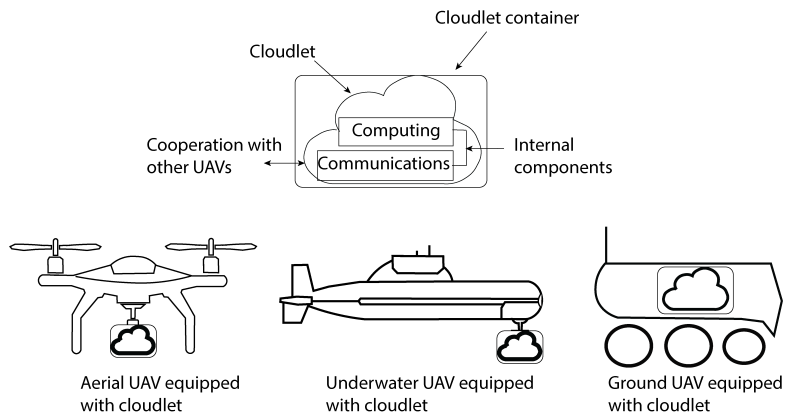} 
\caption{GEESE System Overview}
\label{fig:geesesystem}
\end{figure}

Figure~\ref{fig:geesesystem} describes the overall system design of~\systemname. The core components of~\systemname~consists of commercial-off-the-shelf UAVs, which are augmented with cloudlets. These cloudlets are encased into specially designed containers that allow them to be easily transported and protected from external damage, e.g., water. We rely on off-the-shelf components to ensure a flexible replacement of components, and to enable rapid prototyping of applications. Off-the-shelf components are essential for large-scale adoption of UAVs transporting cloudlets. We next describe the cloudlet component that augments UAVs.

\noindent \textbf{Cloudlet container:} Cloudlet encasing of computing resources is selected based on the transportation modality of each UAV. Containers allow UAVs to augment and replace their computing resources easily just by attaching/detaching them. For aerial UAVs, we developed a lightweight and shockproof encasing made of carton filled with air pillow bags to reduce impact in case of an emergency landing. Similarly, ground UAVs also rely on plastic encasing filled with airbags to fix and protect cloudlets from unexpected turbulence. In the case of underwater UAVs, existing off-the-shelf UAVs solutions include internally, a limited amount of computing resources that are sealed, e.g., PowerVision PowerRay, BlueRobotics ROVs\footnote{https://bluerobotics.com/}. More advanced solutions designed for marine explorations rely on specialized vessels to accommodate lightweight payloads. These vessels are also sealed to protect internal resources from water damage. As an example, eFolaga~\cite{caiti2013mobile} is a torpedo-like underwater UAV that is designed to carry computing resources that aid its navigation and coordination operations. Similarly, other solutions rely on sealed vessels to carry high computing resources that can execute deep learning underwater~\cite{cadena2017modular}. Unlike other work, for underwater UAVs, we use waterproof containers made of glass to resist water pressure. The reason for using glass containers is that besides computing resources, we envisioned underwater UAVs to augment its functionalities with other resources, such as optical sensing and communications. Optical-based functionality can easily work through glass without degrading its performance~\cite{flores2020penguin}. Otherwise, it is necessary to develop new components to handle those functionalities in different containers. We also equipped glass containers with an internal cargo (additional weight) to counter the upward force (upthrust/buoyant force) caused due to internal air residing inside the glass container. By doing this, underwater UAVs can navigate submerged. For all the different UAV modalities, materials of containers were selected in such a way that they do not interfere with the wireless communication technologies embedded in smart and IoT devices. At the same time, the selected encasing was chosen to protect cloudlets from adverse conditions of the environment, while minimizing encasing weight as much as possible.

\noindent \textbf{Cloudlet computing:} Off-the-shelf UAVs tend to have limited computing capabilities. For instance, basic hardware is available in UAVs to execute simple routines to avoid burdening their battery resources, e.g., return to home and navigation tracking. To overcome this limitation,~\systemname~augments processing capabilities of UAVs with cloudlets made from smart and IoT devices~\cite{lagerspetz2019pervasive}. As described previously, these cloudlets are encased into containers that act as a separate component to support computing operations, e.g., image processing, machine and deep learning operations. Multiple devices can be located inside a container to create a resourceful computing infrastructure via collaborative processing~\cite{flores2020cosine}. While there is a maximum amount of computing resources that can be attached to individual UAVs without impacting their navigation and battery resources, multiple UAVs can also cooperate by interconnecting their individual computing resources to form an edge-like computing infrastructure with higher computing capabilities.

\noindent \textbf{Cloudlet communications:} The communication component is also fundamental to reduce the operational computing burden of individual UAVs. By relying on cyber-foraging techniques that exploit communications~\cite{flores2015beyond}, underwater UAVs can augment their individual processing resources and extend their battery life by performing collaborative processing among them. \systemname{} uses off-the-shelf wireless communications to allow devices in proximity accessing cloudlet infrastructure carried by UAVs, and maintaining contact with remote operators. In addition, computing resources of individual UAVs are also interconnected via wireless to augment the pool of processing resources elastically. While~\systemname{}~uses wireless effectively to provide processing services on the edge, communications with underwater UAVs can influence the processing performance of cloudlets. Indeed, underwater UAVs cannot rely on standard wireless interfaces when UAVs are submerged too deep as the wireless signal is absorbed by water (See results in Section~\ref{sec:resultsgeese}). Wireless communication can be used in short distance to enable submerged components to intercommunicate, but not for long-range intercommunication between multiple underwater UAVs that are submerged. Instead, UAVs can rely on wireless effectively just when underwater UAVs are floating on the surface. To counter the limitation of communication when underwater UAVs are submerged,~\systemname{}~requires to implement additional long-range communication underwater using solutions that can work robustly in underwater environments, such as audio, acoustic, optic or electromagnetic communications~\cite{che2010re}. While several works have demonstrated the feasibility of using these technologies for long-range underwater communication, there are no off-the-shelf solutions yet that are mature enough to enable other types of communications underwater. Thus, currently~\systemname{}~relies solely on standard wireless technologies.

\subsection{Implementation}

\systemname~is build on top of commercial off-the-shelf UAVs technologies. Our prototype uses PowerRay, PowerEye, Phantom 3, and DFRobot Romeo V2 UAVs. Each of them is a reasonably priced solution (between $2000-2500$ euros for PowerRay and PowerEye; and $500-1500$ euros for Phantom 3 and DFRobot Romeo V2) to transport cloudlets at low cost. We also used smartphones and micro-controllers as processing units for the cloudlets. In addition, we rely on the embedded WiFi interfaces of these devices to provide communication support (See~\ref{sec:experimental} for a detailed description of the apparatus used in our testbed).

\section{Experimental setup} \label{sec:experimental}

We next demonstrate the feasibility of transporting cloudlets to the edge using UAVs. In our experimental testbeds (Figure~\ref{fig:ExperimentalTestBedDesign}), we quantify the battery consumption required by UAVs to transport cloudlets, and the performance of collaborative processing between UAVs carrying cloudlets.

\subsection{Cloudlet transportation}

Edge computing infrastructure in the proximity of users is important to improve the performance of applications and release constrained devices from processing and battery limitations. Existing solutions to overcome these issues do not provide ubiquitous solutions that fulfill the requirements of users' mobility. As a result, resource-constrained devices have difficulties operating in the wild for longer periods and provide good quality of service to users. In this experiment, we focus on transporting cloudlets using UAVs, such that it is possible to introduce computing support for devices to any environment dynamically and with ease. Unlike static deployments that require continuous maintenance and costly operations, UAVs can foster the adoption of dynamic deployments based on demand and utility cost (pay-as-you-go).      

\noindent \textbf{Apparatus:}  For aerial UAVs, we rely on measurements obtained from two different drones. We consider a PowerVision PowerEye ($4000$ gm), and Phantom-3 drones ($1216$ gm). We use PowerVision PowerRay ($3800$ gm) as an underwater UAV. Likewise, for our ground UAV, we rely on a DFRobot Romeo V2 ($250$ gm). 

\noindent \textbf{Experiments:} We estimate the influence of weight in the operational time of UAVs. To achieve this, we measure the effort induced in UAV battery consumption when UAVs transport cloudlets of different weights. These weights were encased in their respective container depending on each UAV modality. 

\begin{figure*}
    \centering%
     \begin{subfigure}[b]{0.325\textwidth}
    \centering
        \includegraphics[width=.99\columnwidth]{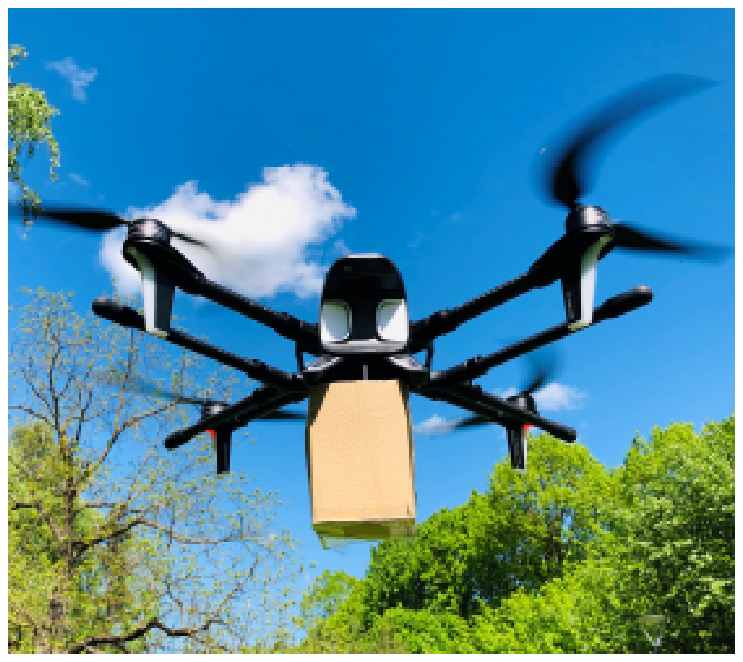}
        \label{fig:flyingdrone-testbed}
                \caption{Aerial UAV cloudlet}
    \end{subfigure}%
     \begin{subfigure}[b]{0.32\textwidth}
    \centering
        \includegraphics[width=.99\columnwidth]{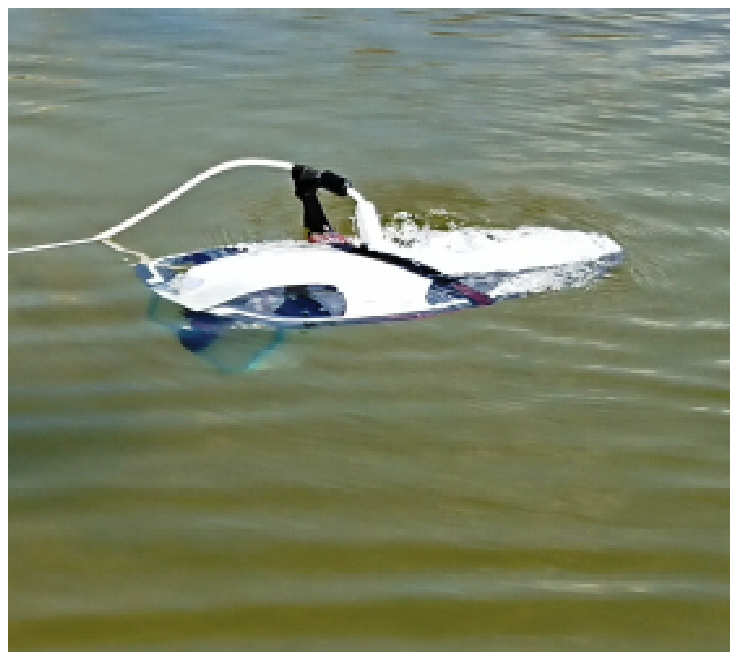}
        \label{fig:underwaterdrone-testbed}
       
         \caption{Underwater UAV cloudlet}
    \end{subfigure} 
    \begin{subfigure}[b]{0.32\textwidth}
    \centering
        \includegraphics[width=.99\columnwidth]{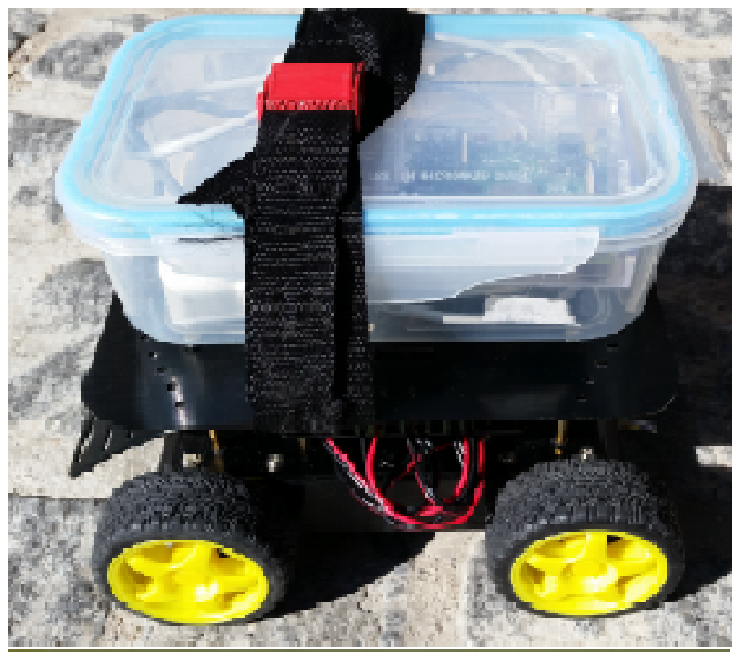}
        \label{fig:grounddrone-testbed}
      
         \caption{Ground UAV cloudlet}
    \end{subfigure}%
     \caption{UAVs with cloudlets and different cloudlet encasing for each UAV modality.}
     \label{fig:ExperimentalTestBedDesign}
\end{figure*}

\begin{figure*}
    \centering%
     \begin{subfigure}[b]{0.33\textwidth}
    \centering
        \includegraphics[width=.99\columnwidth]{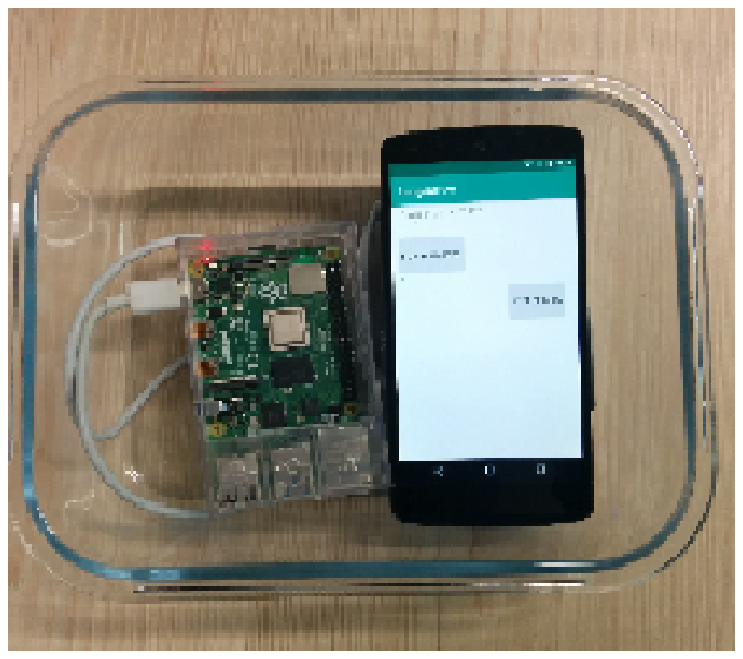}
        \label{fig:underwater-encasing}
       
        \caption{Waterproof encasing}
    \end{subfigure}%
     \begin{subfigure}[b]{0.37\textwidth}
    \centering
        \includegraphics[width=.99\columnwidth]{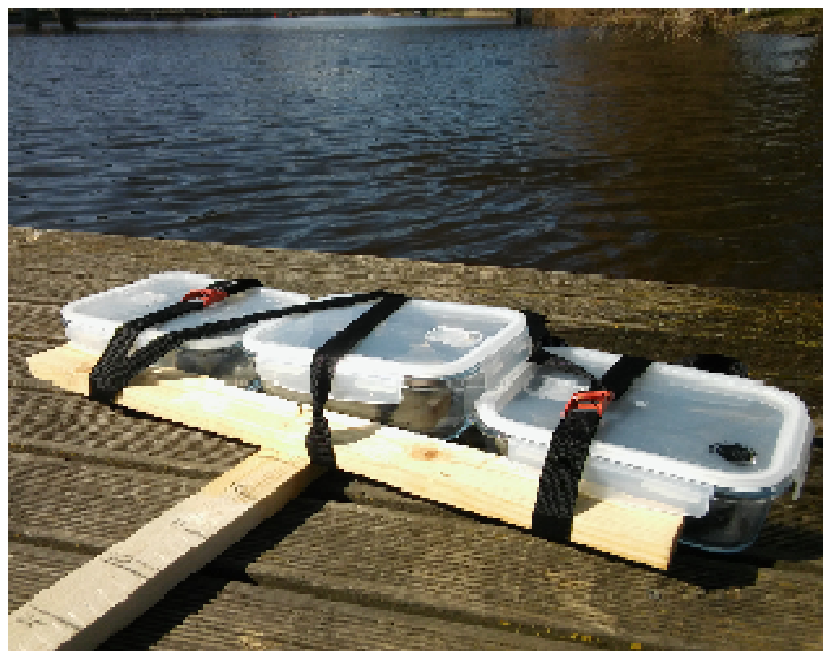}
        \label{fig:collaborative-testbed}
       
         \caption{Cloudlet cooperation}
    \end{subfigure} 
     \caption{Collaborative processing testbed}
     \label{fig:CollaborativeTestBedDesign}
\end{figure*}

\noindent \textbf{Cloudlet weights:} We rely on fixed weights from $100$ gm to $400$ gm to benchmark each UAV. We used the same weights in each drone to make our results comparable between different modalities. For aerial UAVs, we located the different weights in the shockproof casing made of carton and proceed to lift them from the ground (see Figure~\ref{fig:ExperimentalTestBedDesign}a). We ensured that the encased weight was located in the most central region at the bottom of the drone to avoid harming its flying stability. We used weights up to $400$ gm in our testbed. We did not consider heavier weights as those started to induce problems of stability in the drones, which could lead them to crash and damage the equipment seriously. In the case of the underwater UAV, we encased each weight into waterproof containers made of glass (see Figure~\ref{fig:CollaborativeTestBedDesign}a). We focus on analyzing only transportation underwater, not floating on the water surface. Thus, we first neutralize the buoyant force of the waterproof encasing for cloudlets by adding $830$ gm to the glass container. This extra weight is required to eliminate the floating caused due to the upthrust force that takes the glass container to the surface. Existing marine vessels are also designed to compensate this upthrust force. Without this, the underwater UAV drains energy three times faster just for sinking the container and keeping it underwater. Once the floating force was neutralized, we then measure the effort of the underwater drone to carry waterproof cloudlets from $100$ gm to $400$ gm (see Figure~\ref{fig:ExperimentalTestBedDesign}b). Similarly, we encased the same weights ($100$ gm to $400$ gm) in protective plastic containers, and located them in our ground UAV (see Figure~\ref{fig:ExperimentalTestBedDesign}c).

\noindent \textbf{Setup:} To take battery drain measurements, we rely on the Vision+~\footnote{ https://support.eu.powervision.me/support/home/} and DJI GO~\footnote{https://www.dji.com/ee/downloads/products/phantom-3-standard} applications for PowerVision, and Phantom drones, respectively. We install the applications on a mobile phone and configure them to connect to the drones via their base stations. Each application provides real-time information about the resources of the drone, e.g., battery, camera. For underwater UAVs, we measure the energy consumption of the Power Ray drone with the same Vision+ application that we use for the PowerEye drone. The application has a separate interface for the Power Ray drone, which can provide real-time information about the drone's battery status. With each application, we record the battery consumption of each drone while carrying different weights. We record energy drain intervals that depict 10\% of energy consumption. To do this, we monitor how long the UAV can operate while carrying the weight in that battery interval. We consider only intervals above 50\% of energy. For instance, 80\% to 70\%, 90\% to 80\%. We did not consider other intervals as drones have minimal energy policies to operate and to avoid emergency landing due to fast energy drained caused by weight. In the case of the ground UAV, there is no available app to take direct measurements. Thus, to measure energy consumption, we use our own rechargeable battery pack (7000mAh, 6v). First, we charge the battery pack fully with help from the PeakTech Digital Multimeter\footnote{ https://www.peaktech.de/productdetail/kategorie/digital---handmultimeter/produkt/peaktech-3430.html} that is connected serially to the charging circuit and logged the current flow every second. When the Multimeter shows that there is no current flow in the circuit, it means that the battery pack is charged up to its maximum capacity. Then we connect the battery pack to the ground UAV, attached 100gm of weight (cloudlet weight), and start running on the ground until the battery drains completely. Meanwhile, we measure the operational time also. We repeat the same experiment for different cloudlet weights from 100gm to 400gm and measure the time accordingly. We extract from these measurements, the corespondent operational time that is possible in a 10\% energy consumption interval for different weights.

\subsection{Collaborative processing}

A key limitation of edge computing is that it does not scale. Indeed, edge computing resources cannot be augmented dynamically, like its cloud computing counterpart. While we have demonstrated that UAVs can transport cloudlets to the edge, individual UAVs are still constrained to a static amount of computing resources. As a result, collaborative approaches to merge computing resources from different UAVs are critical to augment the amount of computing resources on the edge dynamically. In this experiment, we analyze the performance of collaborative processing to form a dynamic edge computing infrastructure through the cooperation of different types of UAVs, including aerial, underwater, and ground. 

\noindent \textbf{Experiments:} We measure the success rate of job completion from a computational task that is executed in a cooperative manner between multiple UAVs. We analyze the processing performance degradation that can occur when cloudlets intercommunicate in different environments. While underwater UAVs are envisioned to provide connectivity to the cloudlet by floating on the water surface, we analyze different underwater situations that can influence the collaborative processing between UAVs.

\noindent \textbf{Baseline:} As baseline in our experiments, we rely on MobileNet models~\cite{howard2017mobilenets}. We also compared our results with other benchmarks that use MobileNet for collaborative processing with smart devices~\cite{lagerspetz2019pervasive}.

\noindent \textbf{Prototype:}  We implemented a proof-of-concept prototype that follows a master/worker topology, where a worker is an idle device, and the master is an initiator device that triggers the execution. In our prototype (see Figure~\ref{fig:ExperimentalTestBedDesign}c), a single device is elected as a master, while the other remaining devices act as workers. The master is in charge of initiating the task and divide it into jobs based on the number of available devices. Each job is then sent by the master to the workers in a round-robin fashion. Jobs are independent of each other. Thus, the success of the application is measured by the rate of completed jobs. Each worker then processes the job, and returns the task promptly to the master once the task is finished. The master collects the results from the workers, and merges the processed contributions into a single result. Our prototype application is developed in Android version 5.0.1 and implements a Convolutional Neural Network (CNN) model to recognize objects within images. 

\noindent \textbf{Task:} As the experimental task, we consider object recognition from a video-feed. This task was selected as one single device cannot provide suitable response time when processing it. Thus, cooperation among multiple devices is required to improve the response time. We consider a set of 50 images in a resolution of $224x224$, which are collected from the ImageNet~\footnote{http://image-net.org/}. For recognizing objects in the images, we rely on a deep learning implementation that uses TensorFlow~\footnote{https://www.tensorflow.org} with a pre-trained and quantized mobilenet\_v1\_1.0\_224 model. To run our application, we rely on LG Nexus 5 devices.

\noindent \textbf{Setup:} We evaluate the performance of collaborative processing between four devices acting as cloudlets, where one cloudlet is elected as master, and the other three are workers. Since wireless signal communication does not propagate well through water, we analyze the influence of water when performing collaborative processing. This means that we analyze when underwater UAVs engage in cooperative processing with aerial and ground UAVs. To this end, we evaluate two scenarios, 1) the master device is underwater, and workers are above the water surface, and 2) workers are underwater, and the master is above the water surface. For both scenarios, we evaluate different underwater depths.

\section{Results} \label{sec:resultsgeese}

\noindent \textbf{Summary of results}
\begin{itemize}[noitemsep,leftmargin=*] 
    
      \item We equipped off-the-shelf UAVs with portable cloudlets encased in specialized containers. We then measured the energy effort required by UAVs to transport cloudlets to the edge. We found that transporting cloudlets to edge is more energy costly for aerial, followed by underwater, and is least costly for ground UAVs. 
    
    \item Our results also indicate that weight is not a representative factor to define the capacity of computing resources. Indeed, as long as the right combination of aggregated resources is in place, a cloudlet of $300$ gm can have better processing capabilities when compared with a cloudlet of $400$ gm.
    
    \item We found that a single off-the-shelf UAV can transport a cloudlet of  weight up to $400$ gm to the edge. This cloudlet can handle the computational workload of around $900$ concurrent users, and provide an average response time of two seconds when processing 10 images in a resolution of $224x224$.
    
    \item We conduct collaborative processing between different types of UAVs carrying cloudlets. Our results indicate that collaborative processing is even feasible with underwater UAVs. Specifically, we found that underwater UAVs have to be on the surface or at least in a depth no longer than $10$ cm, such that the collaborative processing is stable and without disruptions.

\end{itemize}

\medskip

\begin{figure*}
    \centering%
     \begin{subfigure}[b]{0.5\textwidth}
    \centering
        \includegraphics[width=.99\columnwidth]{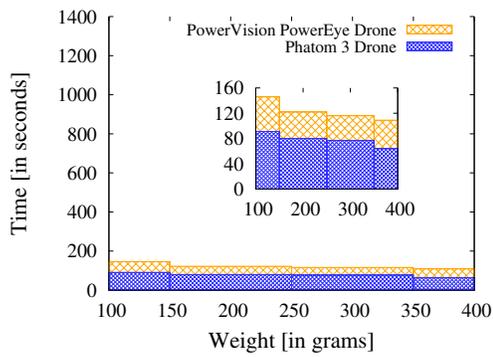}
        \label{fig:flyingdrone-time}
      
        \caption{Aerial UAV}
    \end{subfigure}%
     \begin{subfigure}[b]{0.5\textwidth}
    \centering
        \includegraphics[width=.99\columnwidth]{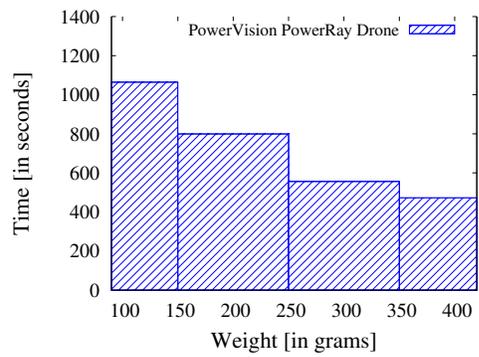}
        \label{fig:underwaterdrone-time}
        
         \caption{Underwater UAV}
    \end{subfigure} \\
    \begin{subfigure}[b]{0.5\textwidth}
    \centering
        \includegraphics[width=.99\columnwidth]{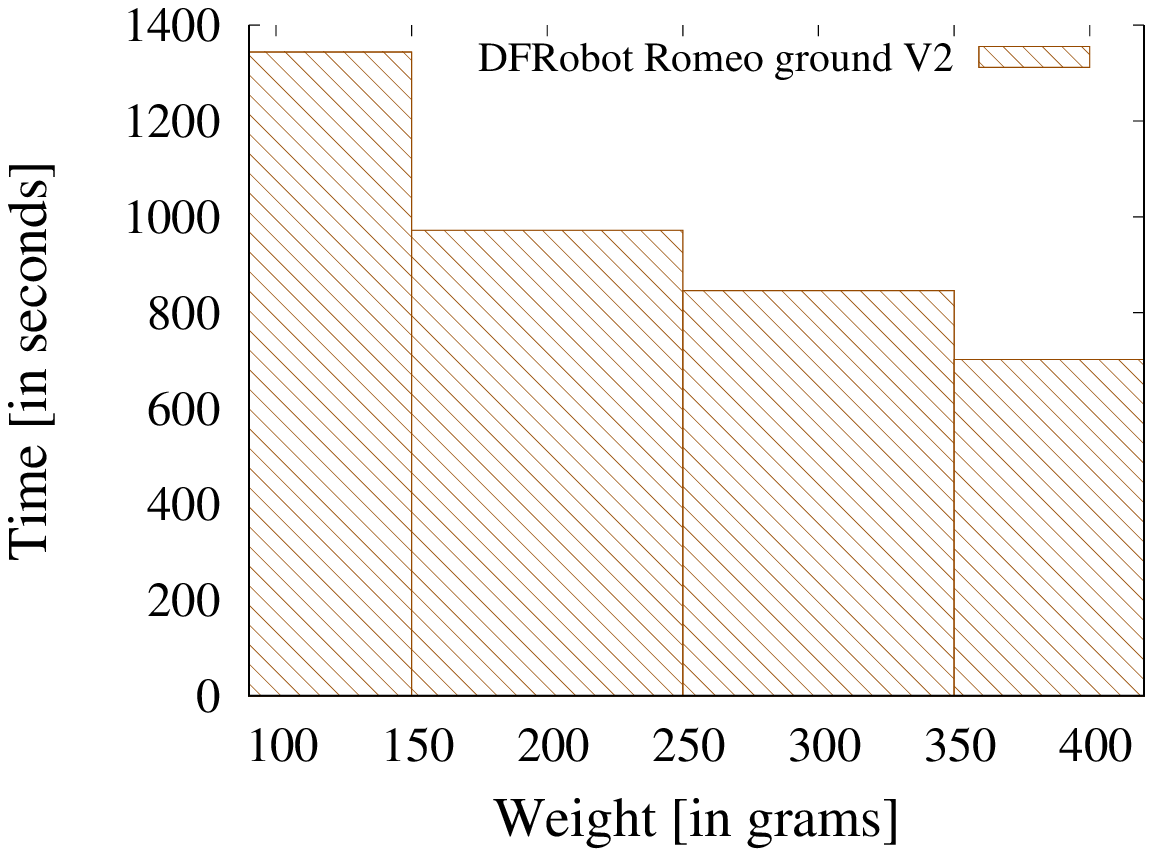}
        \label{fig:grounddrone-time}
      
         \caption{Ground UAV}
    \end{subfigure}%
     \caption{Operational cost in terms of battery consumption for aerial, underwater, and ground UAVs when equipped with cloudlets of different weights.}
     \label{fig:BatteryDroneDelivery}
\end{figure*}

\subsection{Cloudlet transportation performance}

Figure~\ref{fig:BatteryDroneDelivery} shows the results of delivery performance for aerial, underwater, and ground UAVs. From the figure, we can observe that the flying time of UAVs decreases as the cloudlet weight increases for both aerial drones. We can observe a significant difference between cloudlets weights of $100$ gm and $400$ gm. In these cases, the flying time reduces $\approx$ 30\% for PowerEye ($146$ s to $109$ s), and Phantom ($91$ s to $64$ s), respectively. Similarly, when considering the operational time of the PowerRay underwater drone, we can observe that it decreases as the cloudlet weight increases. For instance, when the cloudlet weight is $100$ gm and $400$ gm, the operation time is $1064$ s and $473$ s, respectively. However, the time is significantly high when compared with the aerial drones. This indicates that transporting cloudlets to edge using underwater UAVs is more energy-costlier than aerial UAVs. In the case of underwater UAVs carrying cloudlets, it is also important to highlight that too low or high weight is counterproductive. We find that an empty waterproof encasing itself reduces the navigation time underwater by $70\%$. This is because the UAV needs extra effort to keep the empty waterproof encasing underwater. Conversely, the waterproof cloudlet aids the UAV to save energy when it is on the surface, as it provides extra floating support.

In terms of computing capacity and processing performance, while the battery of UAVs suffers due to the weight attached to the drone, this is compensated by the actual amount of computing resources that can be transported to the edge. To illustrate this, Table~\ref{tbl:cloudletDelivery} shows the actual weight, which is translated into the processing capabilities of potential portable cloudlets. From the table, we can observe the weight carried by the drone (Cloudlet weight), the amount of resources that match that weight (Encased resources), a quantifiable description of the cloudlet resources (Quantifiable description), and the performance of the computing resources that are transported to the edge (Processing performance). We measure the processing performance by analyzing the average response time to process a fixed computing task (average response time when processing $10$ images). Similarly, we measure the capacity performance by analyzing the number of concurrent users that can connect to the resources (No. of concurrent users). From the table, we can observe that a $400$ gm cloudlet can provide computing support to process 10 images in $2.21$ s. In contrast, a $100$ gm cloudlet can provide computing support to process the same 10 images in $50$ s. Naturally, edge computing support powered by cloudlets can be divided into multiple UAVs. For instance, two UAVs that are carrying individually a $200$ gm cloudlet, can match the processing capabilities of a single UAV carrying $400$ gm. In addition, we also present a detail description about the computing specifications of devices used as portable cloudlets in Table~\ref{tbl:cloudletconfig}.

\begin{table*}[t]
\centering
	\begin{adjustwidth}{-1.2cm}{}
		\scalebox{0.70}{ 
		\begin{tabular}{|l|l|l|l|l|}
			\hline
			\minitab{\\\textbf{Cloudlet weight}}  & \minitab{\textbf{Encased resources}}
			&\minitab{\textbf{Quantifiable description}}& \multicolumn{2}{c|} {\textbf{{Processing performance}}} \\ \cline{4-5}
			& & & \textbf{Time to process}  & \textbf{No. of } \\
			& & & \textbf{10 images}&\textbf{concurrent users} \\
			\hline

		\minitab{Category-1\\100 gm } &\minitab{\\{\includegraphics[height=.5cm ]{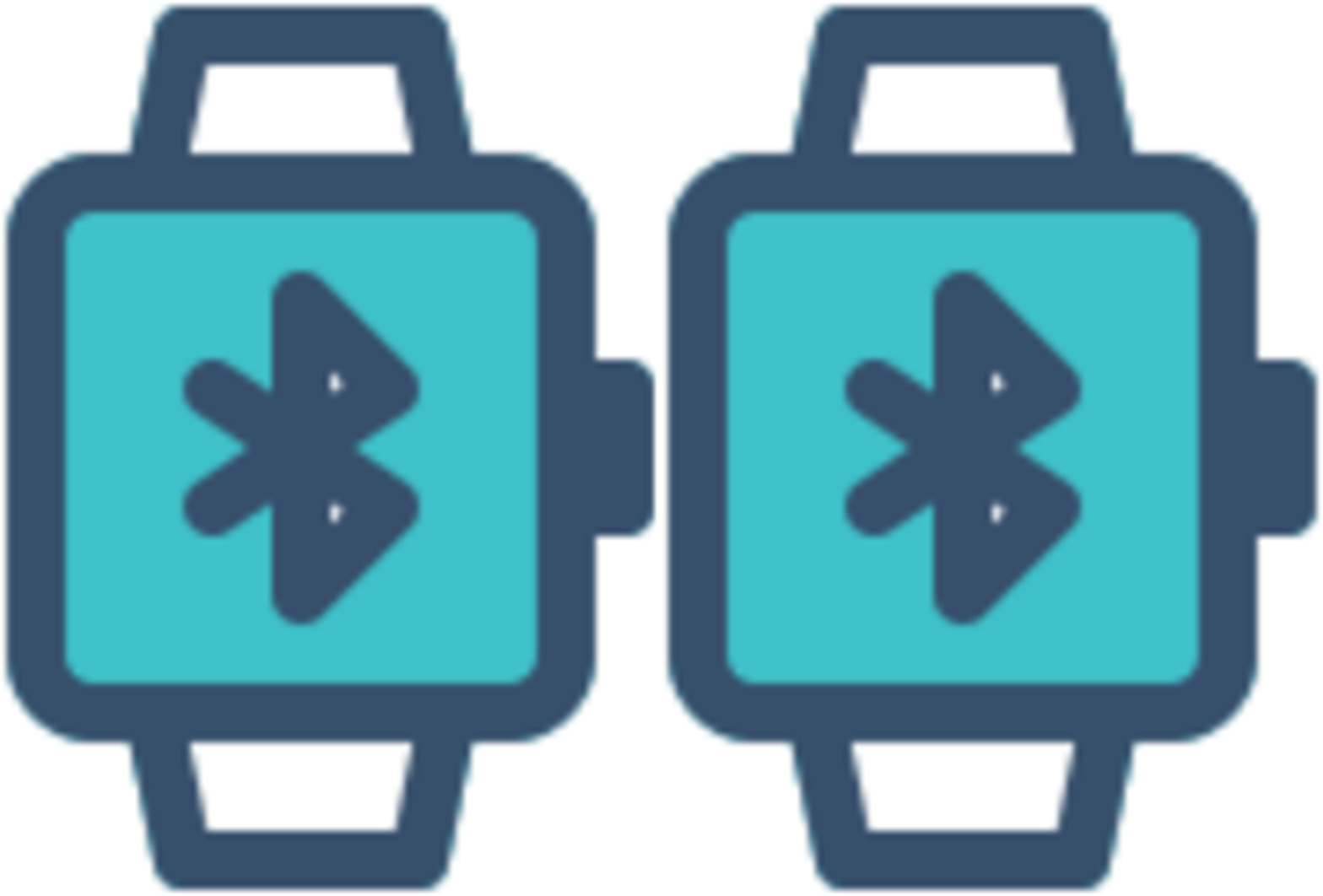}} Type 1} &{\minitab{Samsung Active 2 Smart Watches\\\textbf{(42gmx2=84gm)}}} & 50 seconds& 36 to 40 \\ \hline
						
						 \multirow{3}{*} {\minitab{\\Category-2\\\\200 gm }}	&{\minitab{ \\ \includegraphics[height=.9cm ]{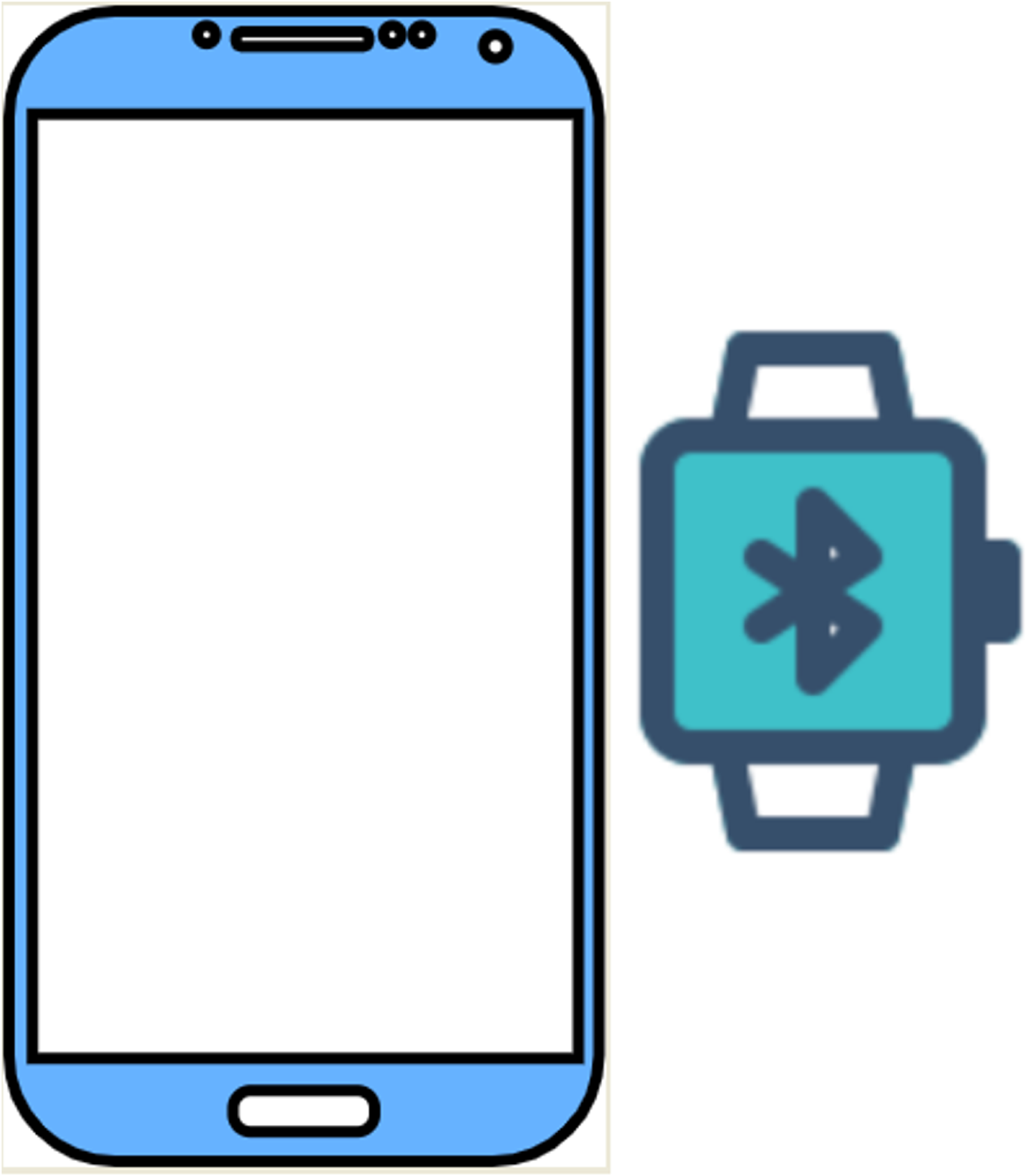} Type 1 }}& {\minitab {LG g4 + Smart Watch\\\textbf{(155gm+42gm=197gm)}}}&37.5 seconds&150 to 170\\ \cline{2-5}
																			&	{\minitab{\\ \includegraphics[height=.9cm ]{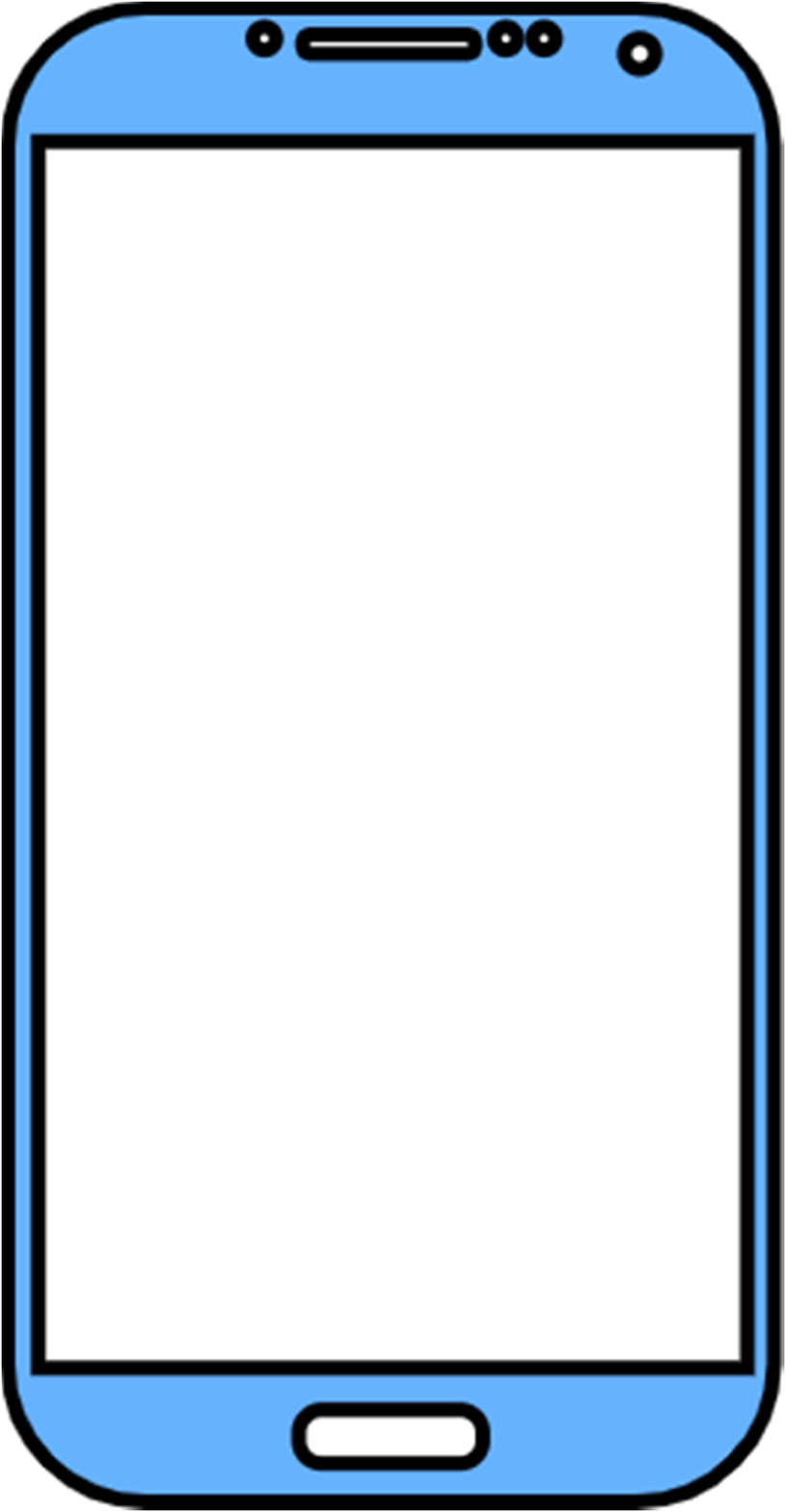} Type 2}}&{\minitab{Sony Xperia\\\textbf{(160gm)}}}&4.46 seconds & 210 to 230\\ \cline{2-5}
																			& {\minitab{\\ \includegraphics[height=.8cm]{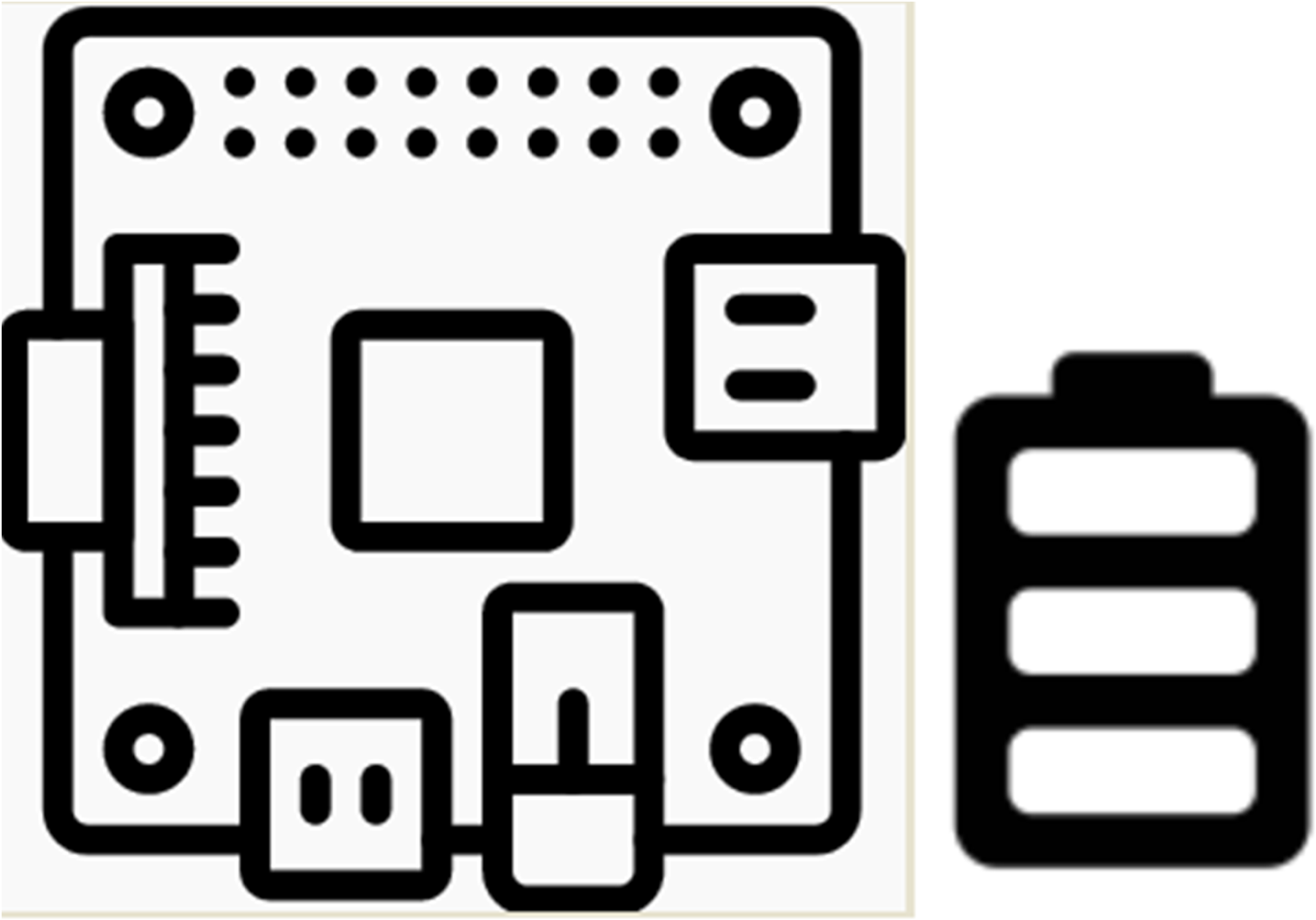} Type 3}}&{\minitab{RP4 with a Battery pack\\\textbf{(48gm+113g=161gm)}}} &58 seconds & 280 to 300 \\ \hline
						
						\multirow{3}{*}{\minitab{\\Category-3\\\\300 gm }}	 &{ \minitab{\\ \includegraphics[height=.9cm ]{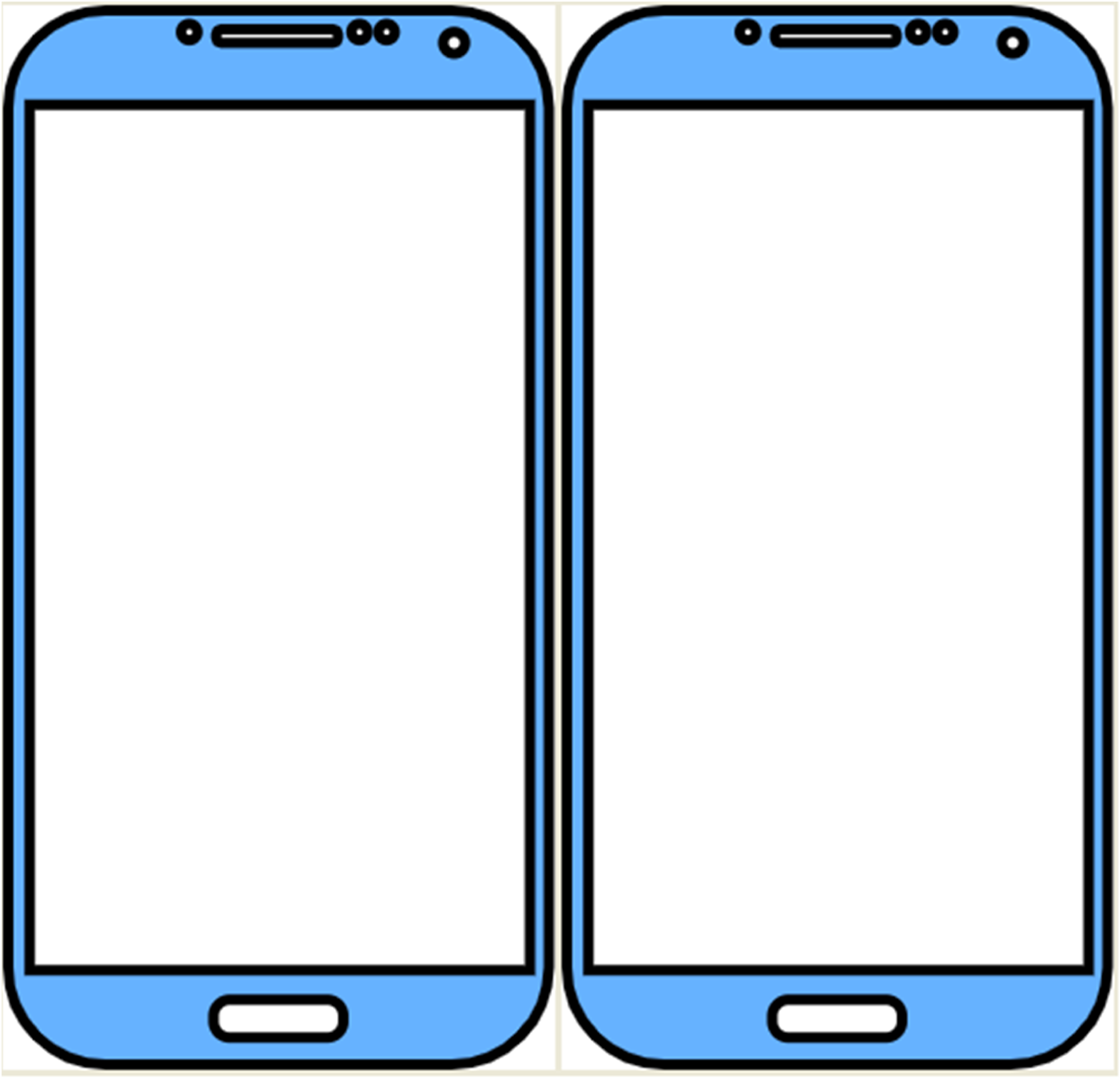} Type 1}}& {\minitab{Samsung Galaxy S5\\\textbf{(140gmx2=280gm)}}} &19 seconds& 300 to 320\\ \cline{2-5}
																			&{\minitab{\\ \includegraphics[height=.9cm ]{graphics/mob2.eps} Type 2}}&{\minitab{Sony Xperia +Samsung Galaxy S5\\\textbf{(160gm+140gm=300gm)}}} &7.4 seconds&360 to 380 \\ \cline{2-5}
																			&{\minitab{\\ \includegraphics[height=.8cm]{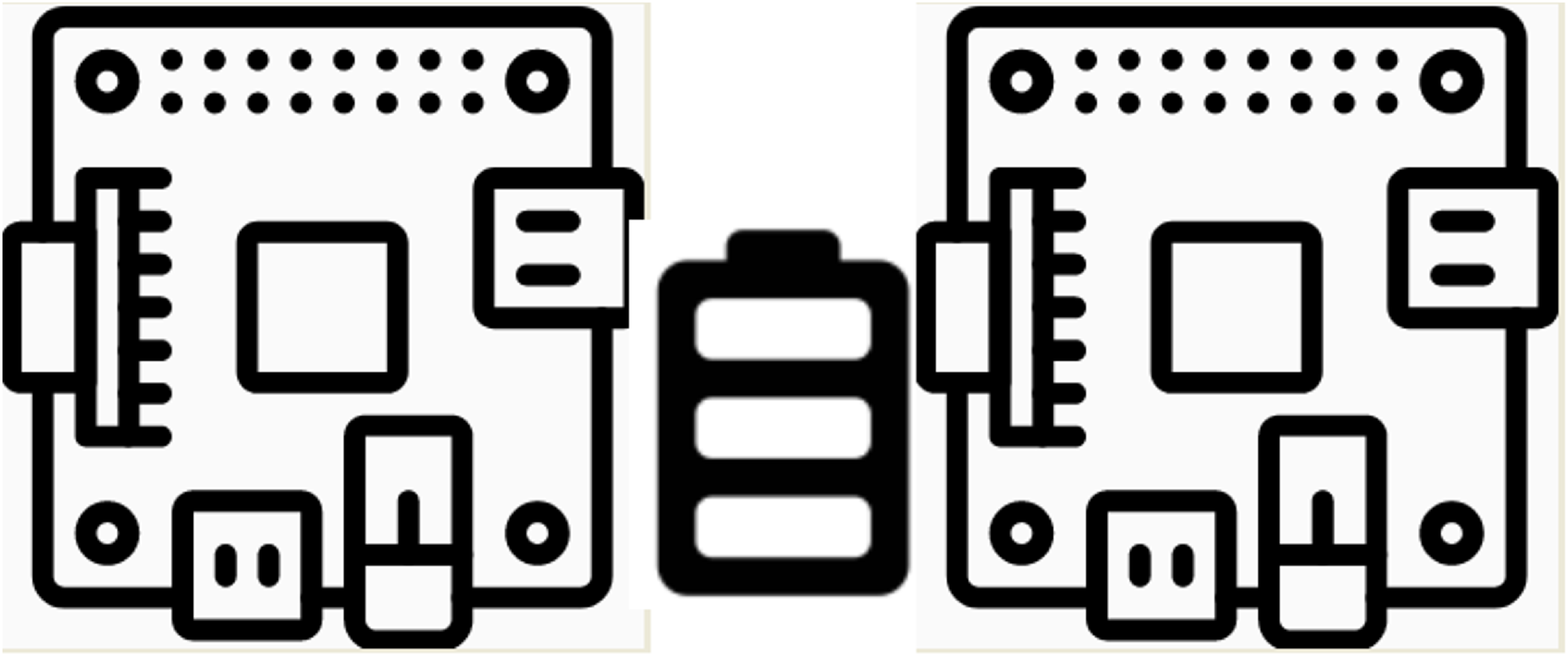} Type 3}}& {\minitab{RP4 with a Battery pack\\\textbf{(48gmx2+113gm=209gm)}}}&29 seconds& 580 to 600 \\	 \hline
						
						\multirow{3}{*} {\minitab{\\Category-4\\\\400 gm }}	&{\minitab{\\ \includegraphics[height=.9cm ]{graphics/mob2.eps} Type 1}}& {\minitab{Sony Xperia\\\textbf{(160gmx2=320gm)}}}  &2.21 seconds& 420 to 440 \\ \cline{2-5}
																			&{ \minitab{\\ \includegraphics[height=.8cm ]{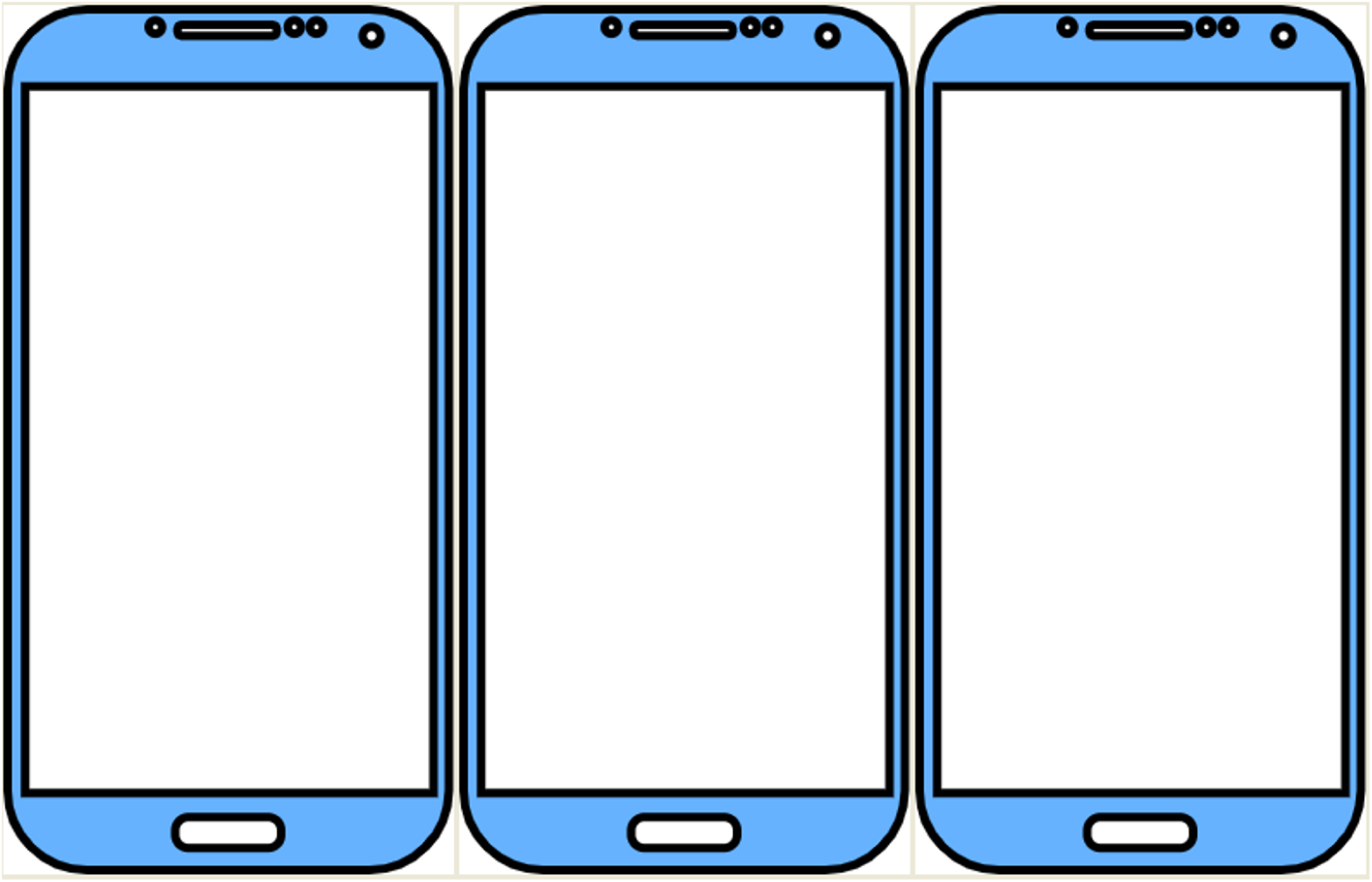} Type 2}}&{\minitab{LG Nexus 5\\\textbf{(130gmx3=390gm)}}}&8 second &400 to 420 \\ \cline{2-5}
																			&{\minitab{\\ \includegraphics[height=.9cm]{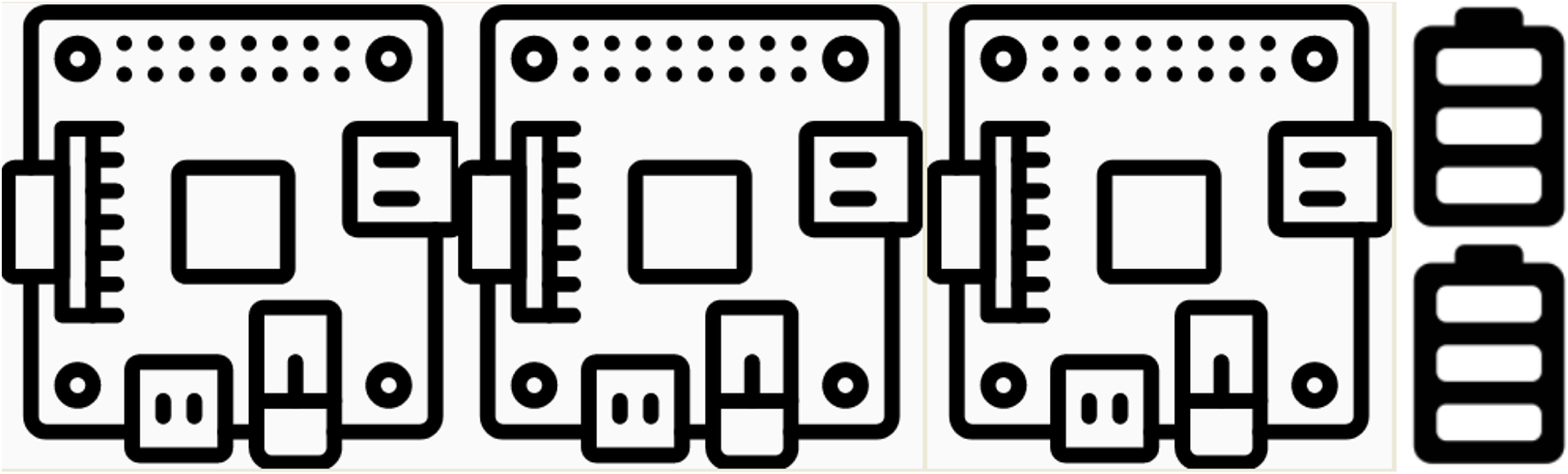} Type 3}}&{\minitab{RP4 with two Battery packs\\\textbf{(48gmx3+113gmx2=370gm)}}} &19.3 seconds  & 880 to 900\\\hline

		\end{tabular}
	}
	\end{adjustwidth}
	\caption{Estimation of computational power in the form of portable cloudlets that can be delivered by UAVs}
	\label{tbl:cloudletDelivery} 
\end{table*}

\begin{table*}[t]
	\centering
	\begin{adjustwidth}{-1.2cm}{}
		\scalebox{0.70}{ 
		\begin{tabular}{|l|l|l|l|}
			\hline
		\textbf{Device} &\textbf{ CPU} & \textbf{GPU} & \textbf{RAM}  \\ \hline	         
	
		{\minitab{Samsung Galaxy Active 2 \\  Smart Watch }}	& Dual-core 1.15 GHz Cortex-A53 &Mali-T720 & 1.5GB 	\\ \hline															
    {\minitab{LG g4\\mobile phone}}& Hexa-core (4x1.4 GHz Cortex-A53 \& 2x1.8 GHz Cortex-A57)& Adreno 418&3GB 	\\ \hline	
		{\minitab{Sony Xperia XZ1 Compact\\mobile phone}} &Octa-core (4x2.45 GHz Kryo \& 4x1.9 GHz Kryo) & Adreno 540& 4GB 	\\ \hline	
				{\minitab{Raspberry Pi 4B	}}& BCM2711, Quad core Cortex-A72 (ARM v8) 1.5GHz& BCM VideoCore VI&4GB  \\ \hline	
			{\minitab{Samsung Galaxy S5\\mobile phone}}	&Quad-core 2.5 GHz Krait 400 & Adreno 330&2GB 	\\ \hline	
			{\minitab{LG Nexus 5\\mobile phone}}	&Quad-core 2.3 GHz Krait 400 & Adreno 330& 2GB	
	\\ \hline					
		\end{tabular}
	
}
	\end{adjustwidth}
	\caption{Computing specifications of devices used as portable cloudlets that are delivered by UAVs.}
	\label{tbl:cloudletconfig} 
\end{table*}

\subsection{Collaborative processing performance}

\begin{figure}
\centering
\includegraphics[width=0.5\textwidth]{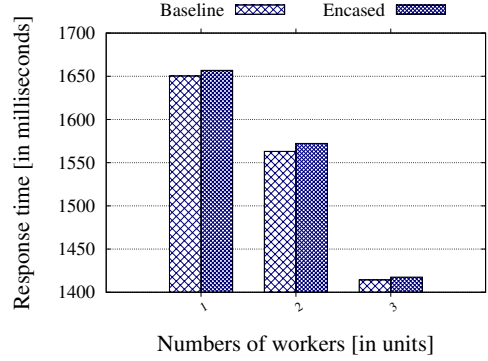}
\caption{Baseline for collaborative processing}
\label{fig:processingUnderwaterBaseline}
\end{figure}

\begin{figure*}[t!]
\centering
  \begin{subfigure}[b]{0.5\columnwidth}
    \centering
    \includegraphics[width=0.99\columnwidth]{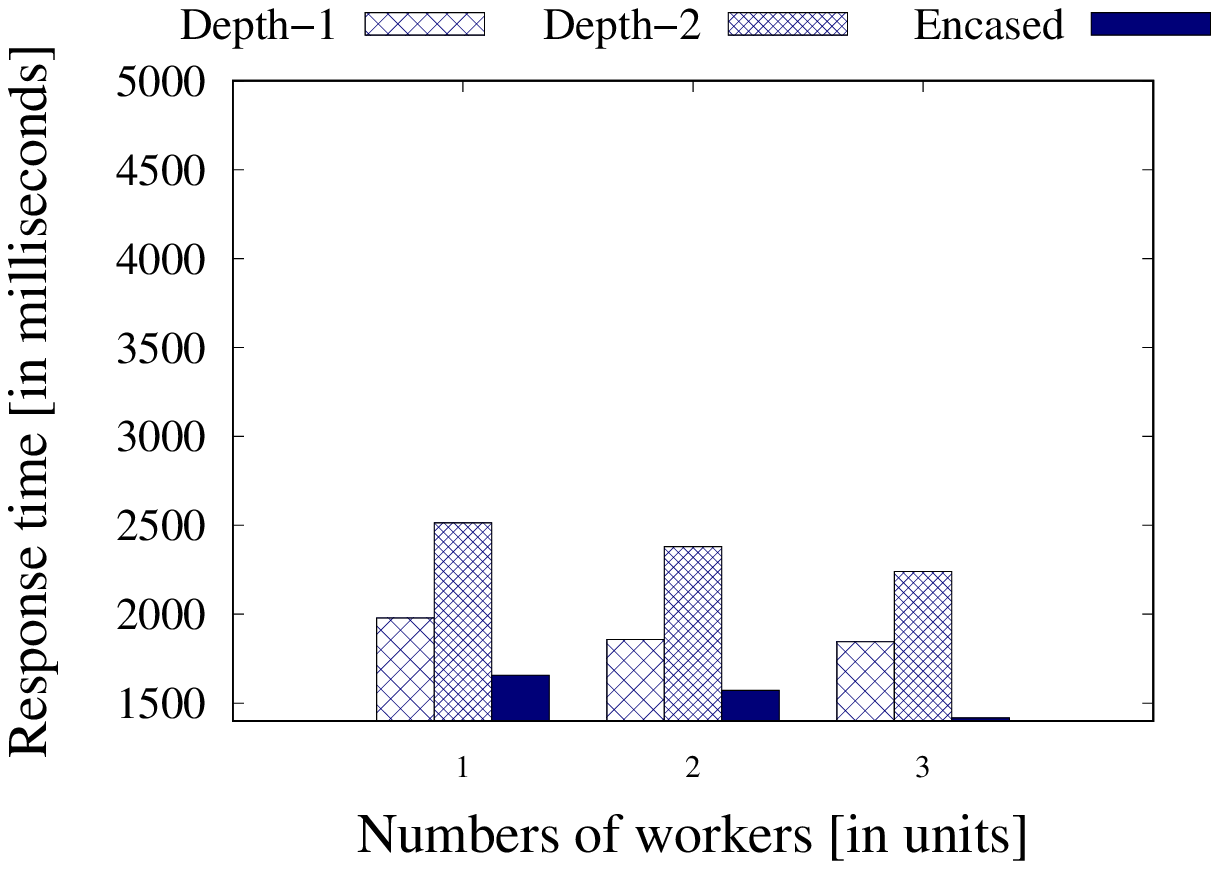}
	\caption{}
	\label{fig:MasterAirWorkersUnderwater}
  \end{subfigure}%
    \begin{subfigure}[b]{0.5\columnwidth}
    \centering
    \includegraphics[width=0.99\columnwidth]{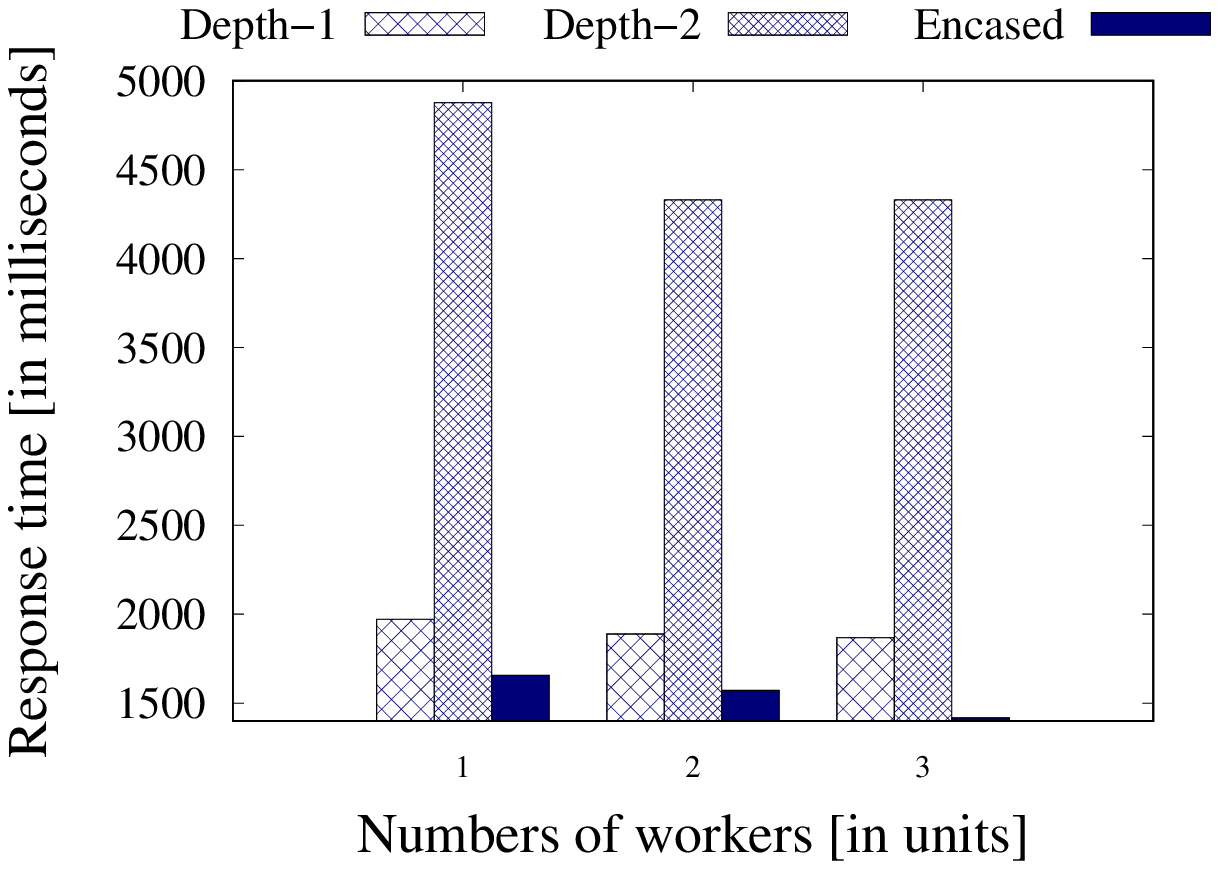}
	\caption{}
	\label{fig:WorkersAirMasterUnderwater}
  \end{subfigure}%
  \caption{Collaborative processing results, (a) Master device above the water surface and worker devices underwater at different depths, (b) Worker devices device above the water surface and master device underwater at different depths}
  \label{fig:collaborativeProcessingResults}
\end{figure*}

Figure~\ref{fig:processingUnderwaterBaseline} shows the performance results of collaborative processing. We include the results of our baseline for comparison purposes. From these results, we can observe that when the number of available devices increases, the processing time gets shorter. This pattern has been reported by other works~\cite{lagerspetz2019pervasive}. Thus, it validates the correctness of our testbed. We also include the results of the baseline when the smart devices are encased into waterproof containers (encased). As we can observe, the containers (glass) do not significantly influence the communication between devices for collaborative processing. For both, baseline and encased setup (with no water), the success rate of job completion is 100\%. In contrast, when the waterproof setup is taken to underwater, the success rate of job completion starts to reduce as the distance between the cloudlets increases. Figure~\ref{fig:collaborativeProcessingResults} shows the performance results of collaborative processing underwater at different underwater depths. The first depth considers distances below the water surface between $5-8$cm. Similarly, depth-2 considers distances below the water surface between $10-12$cm. We did not consider deeper distances as the collaborative processing degrade to unacceptable levels. Figure~\ref{fig:collaborativeProcessingResults}a shows the results of setup, when the master is above the surface, and the workers are underwater. We also included the encased baseline results for comparison purposes. From these results, we can observe that the response time degrades significantly when changing from depth-1 to depth-2. Indeed, while the response time in depth-1 degraded, the success rate of job completion achieved 100\%. Conversely, the success rate decays about $30$\% at depth-2. Similarly,  Figure~\ref{fig:collaborativeProcessingResults}b shows the results of the setup when the workers are above the water surface, and the master in underwater. We can observe that when the master is underwater, the setup suffers more from water impact. From the results, we can observe that when the master is at depth-1, the response time degrades, and also the success rate of job completion decays about $10$\%. Moreover, when the master is taken to depth-2, the success rate decays largely about $38$\%, and the response time degrades 3x times. 

\section{Edge delivery model} 
Since deciding the amount of UAVs that need to be used for cloudlet delivery is important to use~\systemname~optimally, in this section, we introduce an edge delivery model that allows~\systemname~to schedule the delivery of computing resources to the edge. In the following section, we describe the model and its implementation. We rely on the results from the previous section, to analyze the transportation of cloudlets in different situations.

\subsection{Optimal cloudlets to edge}
Consider a job which is requested for a specific workload ($W$) and average response time ($\uptau$), the objective of the model is to minimize the allocation of the number of UAVs ($U_{AV}$) to satisfy the request. The amount of UAVs is estimated using Linear Programming model \cite{taha2013operations}, which includes the following parameters:

\begin{enumerate}
\item ($W$): Workload is defined by the number of users a particular request has asked for.
\item ($\uptau$): Defines the minimum response time that needs to be provided for a particular workload.
\item $C$: Represents the set of available cloudlets that can be placed on an UAV. An \textit{Instance} of a particular cloudlet could be two smartwatches of Samsung used as a cloudlet phones (see Table \ref{tbl:cloudletDelivery}, row 1). Each cloudlet has three characteristics, i) weight of the cloudlet, ii) the amount of workload it can satisfy, and iii) the response time it can provide for a specific workload.
\item \textit{$\top_r$} signify the round trip time that a particular request would need. For example, if a request has been made to the service provider first for a location $l_{1}$ and later for $l_2$, then $\top_r$ is the amount of time a UAV should last for starting its journey from the source location to $l_{1}$, then to $l_{2}$, and finally coming back to the source location. $\top_r$ also includes the time needed for a UAV to stay at locations $l_{1}$ and $l_{2}$.

\end{enumerate}

To optimize the number of UAVs to be deployed, we need to allocate various cloudlets across multiple UAVs such that the overall cloudlets are able to provide at least response time $\uptau$ for a particular workload $W$. The model tries to minimize the cost by selecting different types of UAVs and the cloudlets in such a way that it satisfies the requested job. The objective function is defined as the sum of costs associated across all the selected UAVs and cloudlets.

\begin{equation}
    Min \left(\sum\limits_{i=1}^n \alpha_i U_i+\sum\limits_{j=1}^m \beta_j C_j\right)
\end{equation}

The model consists of the following constraints:
\begin{equation}
    \sum\limits_{j=1}^m C_j w_j \geq W_{r}
\end{equation}

The workload constraint states that the sum of all the workloads across all the cloudlets $C_j$ $\in$ $C$ must be enough to satisfy the workload $W_r$, the workload requested for a particular job.

\begin{equation}
    \frac{1}{U}\sum\limits_{i=1}^n U_i \uptau \leq \uptau_{min}
\end{equation}

The response time constraint states that the average sum of all the cloudlets 
places across all the UAVs must be enough to satisfy the minimum response time requested for $\uptau_{min}$, where $U$ represents the set of UAVs being deployed for a specific job.

In addition, each UAV should be able to reach back to the source location before the consumption of the complete battery. This constraint is represented using the total round trip time for each $U_i$ $\in$ U,  and is represented as the follows:

\begin{equation}
   \sum\limits_{k=1}^l U_i \top_i \leq \top_{req} , \forall U_i \in U
\end{equation}

The round trip time constraint states that for a particular $U_i$ $\in$ $U$, the batteries of this UAV should be able to travel to multiple locations if required and serve the workloads at specific locations and then able to return the source back before the consumption of the battery of UAV and the total round trip time required to serve the requests be $\top_{req}$

\subsection{Evaluation and results}

\noindent \textbf{Implementation:} Our model is implemented in R framework using the~\textit{lpSolveAPI} package. The overall model contains $\approx$  100 LoC and can be easily executed by constrained devices without inducing resource-intensive processing.~\systemname~model is the interface for interaction between UAVs and operators that assemble the cloudlet payload of UAVs, such that drones can deliver the computing resources on the edge. While it is possible to envision a complete autonomous end-to-end process, our current~\systemname~system provides the first step towards it.

\begin{figure}
\centering
\includegraphics[width=0.85\textwidth]{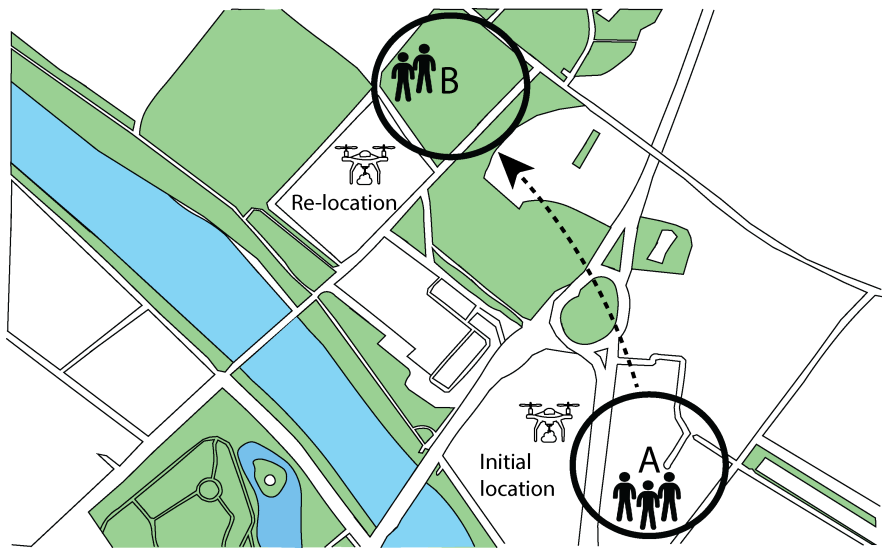} 
\caption{Transportation of cloudlets to the edge to handle the workload}
\label{fig:usecase}
\end{figure}

\noindent \textbf{Use case:} To test our model, we consider a simple use case of a mobile crowd that moves between two different locations. The use case is representative of a common protest that has a start and end points. The situation is presented in Figure~\ref{fig:usecase}. From the figure, we can observe the two locations A and B. The mobile crowd consists of around $1500$ users. Initially, the mobile crowd will gather at location A, and after an hour, the mobile crowd will move to location B, where it will dissolve after two hours. To support the activities of end applications in the two locations, it is necessary to deploy edge computing infrastructure. In our use case, a remote operator decides using~\systemname~delivery model, the amount of UAVs that are required to transport an edge computing infrastructure that can handle the workloads of the locations. Thus, the remote operator is the one that feds the parameters of the delivery model based on visual inspections of the locations using a remote interface. These types of approaches have been widely adopted by other systems~\cite{chen2007human}, and can become autonomous simply by integrating crowd counting approaches, e.g., camera counting~\cite{ryan2014scene}.

\noindent \textbf{Setup:} As cloudlets that can be transported to the edge. We consider the cloudlets described in Table~\ref{tbl:cloudletDelivery} in Section~\ref{sec:resultsgeese}. From the table, we can observe four different categories of cloudlets based on weight. For instance, Category-1 = cloudlets of $100$ gm, Category-2 = cloudlets of $200$ gm and so on. Each category has different computing resource types that can be encased. For instance, Category-2 has three types. Each type has specific computing capabilities and capacity to handle the workload. For example, Category-2, type 1 cloudlet is capable of handling up to 170 users concurrently. Similarly, Category-2, type 2 cloudlet is capable to handle up to 230 users concurrently. We also assigned weights to each UAV modality to depict that the cost of transporting a cloudlet by each modality is different. From our results, we know that aerial UAVs are the costlier, whereas ground UAVs are the cheapest. In addition, we also assume that the remote operator makes the following considerations for each location. For location A, it is possible to use any of the three modalities. In contrast, for location B, it is possible to rely only on aerial and ground UAVs. Lastly, the computing infrastructure to be transported to the edge should handle a minimum workload $1500$ users with a maximum response time of two seconds.

\noindent \textbf{Results:} After our delivery model is fed with all the required parameters, we found that for location A, the optimal amount of cloudlets to be delivered includes two ground UAVs of Category-3 type3, and one underwater UAV Category-2 type2. Similarly, for location B, the required amount of cloudlets to be transported included, two ground UAVs Category-3 type 2, and one ground UAV Category-4 type2. We also evaluate the transportation of cloudlets when locations are combined. This means that cloudlets are first transported to location A, and then re-allocated to location B. In this situation, our model recommends to relying solely on ground UAVs. More specifically, cloudlets to be delivered, include, four ground UAVs in total, two Category-3 type2, and two Category-4 type1. It is interesting to notice that aerial UAVs are not selected as the operational cost is very high when compared with other UAV modalities. Naturally, there are situations where only aerial UAVs can be used. Thus, these results are just tied to this particular case.

\section{Discussion}

Naturally, there is room for further work and improvements. We discuss a few points here.

\noindent \textbf{Implications for edge computing deployments:} Existing edge computing deployments are scarce and far to be dense in the wild. Thus, edge computing is not ubiquitous, and applications cannot rely on it in a continuous manner. While existing works have explored the placement of physical edge servers in the wild to cover highly populated areas under many constraints, e.g., budget and cost, these deployments are powered by cloud and cellular vendors. Thus, end users do not have any control nor can manage the placement of edge servers. Moreover, third party companies that want to provide services on the edge, they can lease these deployments just to push software services into the edge, such that these services can be provisioned to users. In this context, the implications of our work are multiple. First, UAVs can be used to improve the coverage of edge computing deployments in areas where currently there is no infrastructure. Second, UAVs can complement current edge deployments and help them to manage the dynamic workload of users. Third, UAVs can allow third parties to have physical edge infrastructure without constraints. For instance, an app development company that has popular app games can schedule UAVs to certain crowded areas to improve the quality of experience of players. Lastly, UAVs can integrate computing resources to any environment, and even in adverse contexts, e.g., underwater, jungle.

\noindent \textbf{UAV delivery:} While we are aware that more resourceful UAVs exist for delivering goods and services, e.g., Amazon Prime Air, and Starship, among others, we develop our system with off-the-shelf technologies to enable faster prototyping. Moreover, off-the-shelf technologies are key to enable solutions that can be build based on demand dynamically. In addition,  off-the-shelf components are important to facilitate deployments at large-scale. Naturally, our results can be easily extrapolated and generalized to other types of delivery drones.

\noindent \textbf{Room for improvement:} In our work, we explored the encasing of cloudlets, such that these can be embedded to UAVs. While we rely on encasing materials to protect the cloudlets, e.g., waterproof glass containers, other materials could also be considered. We are interested in exploring new encasing materials that not just allow the transportation of computational resources, but also allows transportation of other resources, such as sensing resources. Naturally, the selection of proper encasing requires further evaluations and to take into consideration new requirements. For instance, the material to encase sensors does not have to degrade the sensing performance of the sensors. We are also interested in exploring new materials which are lightweight, such that the transportation of resources to the edge can be maximized. Lastly, we are also interested in exploring further other approaches to counter the upthrust force for underwater containers. Additional cargo to make underwater containers stable and lightweight are critical points to design off-the-shelf components for underwater UAVs.

\noindent \textbf{Other factors:} We demonstrate in our work that UAVs can be equipped with cloudlets to deliver computational resources on the edge. While we took into consideration critical issues about performance, resource allocation, cloudlet weight, and computational capacity of cloudlets, several other factors can influence the adoption of UAVs in the wild. For instance, operational regulations for UAVs are based on restricted areas and protected regions defined by governmental and military institutions from different countries. Another issue is the integration of UAVs with existing legacy technologies, and user interfaces to manage and schedule easily the delivery operations of UAVs. Moreover, from a user point of view, an additional factor is the perception of users towards UAVs operating autonomously and invading public areas for landing and anchoring. 

\noindent \textbf{Edge workload and UAV assistance:} We demonstrate that it is possible to estimate the amount of UAVs needed to handle a particular workload of users on the edge. While edge workload information is easy to extrapolate from cellular operator traces and crowdsensing apps, we can also rely on the cameras integrated within UAVs to perform crowd counting. By doing this, it is possible to request UAVs when there is a high arrival rate of users in a location. Thus, besides the UAVs that deliver cloudlets, we envision additional UAVs that can assist in the process of estimating and monitoring different properties of the edge workload, e.g., group mobility and number of users; such that new UAVs can be scheduled, reallocated or replaced based on demand. 

\noindent \textbf{On cloudlet transportation optimization:} While we present an adaptive model to transport cloudlets to the edge, several other factors can be considered to optimize further the transportation and placement of UAVs. For instance, in our model, we do not take into consideration transportation trajectories and events that can prevent drones from delivering cloudlets, e.g., close and alternative routes. Another factor that can influence the transportation of cloudlets is weather conditions. The transportation time of cloudlets can differ significantly or become unpractical, e.g., storms, snowfalls, and floodings.

\noindent \textbf{Recycling opportunities for e-waste:} Electronic waste from smart devices is a global concern as it pollutes natural ecosystems and fosters climate change. In our work, we demonstrate that UAVs can carry cloudlets made from smart and IoT devices. We envision that computing resources from old phones can be recycled to create portable computing racks, which then can be transported or deployed on the edge to provide public services to users. For instance, a video streaming service for tourists about the sightseeing places in a city.

\noindent \textbf{Composition of computing resources:} While it is possible to collect performance metrics about the execution of applications in different underlying hardware (using computing benchmarks), it is difficult to generalize a solution to obtain an optimal composition of multiple computing resources. Despite the availability of several methods for optimization, e.g., heuristics~\cite{hagras2003simple}, the rapid increase of processing capabilities of smart and IoT devices, makes it difficult to explore all possible combinations of devices and available models. In our work, we relied on a computing benchmark to demonstrate that it is possible to execute different types of applications in portable cloudlets. The optimal composition of those taking into consideration all the different hardware properties is out of the scope of this work.

\noindent \textbf{Natural hazards and malfunctions:} Invading natural ecosystems with technology that is unknown to wildlife has led to numerous attacks and destruction towards UAV deployments. One way to overcome this problem is to pre-analyze the predominant species in the target area of the deployment and adapt the UAV behavior accordingly to avoid interrupting wildlife routines. For instance, drones can be turned off during the night to avoid colliding with species that hunt during that period. Another solution is to apply camouflage techniques to transform the visual appearance of UAVs to blend with the environment. For instance, UAVs can be designed with fish or bird appearance to become (almost) unnoticeable to others. 

\noindent \textbf{Towards autonomous fog:} One important insight that is demonstrated in our work is that computing resources are not just descending to the edge to be in proximity of users, but it is also possible to grant mobility capabilities to the edge through UAVs, such that it can adapt to users mobility characteristics. In this manner, UAVs can adapt the provisioning on edge computing resources, such that the quality of experience and service that is perceived by users is consistent. As a result, our work provides a further step to achieve a computing infrastructure on the edge with fog-like properties. 
\section{Summary and Conclusions}

In this paper, we contribute by designing GEESE, a novel UAV-based system that can transport cloudlets to the edge using different UAV modalities. We analyze the effort required by UAVs to deliver cloudlets of different weights and sizes. We also analyze the challenges of using aerial, ground and underwater UAVs for cloudlet transporation. In addition, we perform rigorous resource allocation experiments to estimate the amount of UAVs required to handle a dynamic workload of users on the edge. Our results indicate the UAVs provide more flexibility and consistency to provision edge computing services when compared to other work that focus on static edge server placement. Indeed, we demonstrate that UAVs can be easily reallocated, replaced, and aggregated on-demand based on workload requirements. We also present the implications of our work for edge server deployments and highlight the importance of UAVs to achieve scalable services on the edge.


\section{References}

\begin{spacing}{0.1}
\bibliographystyle{elsarticle-num}
{\scriptsize 
\bibliography{UAV-MEC}}
\end{spacing}

\end{document}